\documentclass[12pt]{article}
\usepackage[dvips,colorlinks=true,bookmarks=true]{hyperref}
\usepackage{amsmath,amsxtra}
\usepackage{amssymb,amsthm}
\usepackage[english]{babel}
\usepackage[textwidth=18cm,textheight=23cm]{geometry}
\usepackage{graphicx}
\usepackage{psfrag}
\usepackage{array}
\newcommand{\C}{\mathbb{C}}
\newcommand{\s}{\mathbb{S}}
\newcommand{\R}{\mathbb{R}}

\newcommand{\g}{\mathfrak{g}}
\newcommand{\n}{\mathfrak{n}}
\renewcommand{\r}{\mathfrak{r}}
\renewcommand{\d}{\mathrm{d}}\newcommand{\I}{\mathrm{i}}
\newcommand{\bt}{\boldsymbol{t}}

\newcommand{\ad}{\operatorname{ad}} \newcommand{\Ad}{\operatorname{Ad}}
\newtheorem{pro}{Proposition}

\newcommand{\Exp}[1]{\operatorname{e}^{#1}}
\newcommand{\pde}[2]{\dfrac{\partial #1}{\partial #2}}

\newtheorem{teh}{Theorem}

\newtheorem{cor}{Corollary}

\newcommand{\be}{\begin{equation}}
\newcommand{\ee}{\end{equation}}

\begin{document}

\title{On the Whitham hierarchy:\\ dressing scheme, string equations and additional symmetries}

\author{Manuel Ma\~{n}as\\ Departamento de F\'{\i}sica Te\'{o}rica II, Universidad Complutense\\ 28040-Madrid, Spain\\
email: manuel.manas@fis.ucm.es\\ \\
Elena Medina\\ Departamento de Matem\'{a}ticas, Universidad de C\'{a}diz\\
11520-Puerto Real, C\'{a}diz, Spain\\ email: elena.medina@uca.es\\
Luis Mart\'{\i}nez Alonso\\ Departamento de F\'{\i}sica Te\'{o}rica II, Universidad Complutense\\
28040-Madrid, Spain\\ email: luism@fis.ucm.es }

\date{1991 MSC: 58B20}

\maketitle

\abstract{A new description of the universal Whitham hierarchy in
terms of a factorization problem in the Lie group of canonical
transformations is provided. This scheme allows us to give a natural
description of dressing transformations,  string equations and
additional symmetries for the Whitham hierarchy. We show  how to
dress any given solution and prove that any solution of the
hierarchy may be undressed, and therefore comes  from a
factorization of a canonical transformation. A particulary important
function, related to the $\tau$-function, appears as a potential of
the hierarchy. We introduce a class of string equations which
extends and contains previous classes of string equations considered
by Krichever and by Takasaki and Takebe. The scheme is also applied
for an convenient derivation of additional symmetries. Moreover, new
functional symmetries of the Zakharov extension of the Benney gas
equations are given and the action of additional symmetries over the
potential  in terms of linear PDEs is characterized.}

\section{Introduction}

 Dispersionless integrable models, see  \cite{h1},\cite{h2}, \cite{benney} and
 \cite{krich1},
appear in the analysis of various problems in physics and applied
mathematics from the theory of quantum fields, see \cite{krichever}
and \cite{kod}, to the theory of conformal and quasiconformal maps
on the complex plane, see \cite{gib}-\cite{kon1}. The new millennium
brought new applications of these models, see
\cite{zab1}-\cite{zab7}, in different areas, as for
 example integrable deformations of conformal maps  e interfacial
processes.

The Krichever's \emph{universal Whitham hierarchies}, see
\cite{krich1}-\cite{krichever}, are the integrable systems involved
in these applications. These hierarchies include as particular cases
the dispersionless KP, dispersionless modified KP and dispersionless
Toda hierarchies, see \cite{zab1}-\cite{zab4} and
\cite{kaz}-\cite{24}. The role of \emph{twistor} or \emph{string}
equations for studying dispersionless integrable models was
emphasized by Takasaki and Takebe in \cite{24}-\cite{21}. Solutions
of these \emph{string equations}  have attractive mathematical
properties as well as interesting physical meaning.

The objective of this paper is to formulate the factorization
problem for the zero genus Whitham hierarchy within the context of
Lie groups of symplectic transformations, and to give a natural and
general formalism for string equations and additional symmetries. In
particular, we characterize an special class of string equations,
related to a Virasoro algebra. It turns out that this class
determines not only the solutions of the algebraic orbits of the
Whitham hierarchy \cite{krichever}  but also the solutions arising
in the above mentioned applications of dispersionless integrable
models \cite{mel2}-\cite{mel3}.

The layout of the paper is as follows. In \S 2 we introduce the Lie
algebraic splitting for Hamiltonian vector fields and the
corresponding factorization problem for canonical transformations.
Then, in \S 3 we show how deformations of the factorization problem
of canonical transformations lead to solutions of the Whitham
hierarchy. We remark a particular system of equations within the
hierarchy: The Boyer--Finley--Benney equations, which extend the
Boyer--Finley and the Benney equations, respectively. Here we also
introduce a potential function of the hierarchy from which all the
fields of the hierarchy are gotten by appropriate derivations. In a
forthcoming paper \cite{futuro} we show that this function is the
$x$-derivative of  $-\log\tau$, where $\tau$ is the $\tau$-function
of the hierarchy. We proof that any solution of the Whitham
hierarchy may be obtained from a factorization problem; i.e. it may
be undressed. To conclude the section we extend the factorization
scheme to get the dressing of any given solution of the Whitham
hierarchy. In  section 4 we consider the string equations  in the
context of the factorization problem. For that aim we introduce the
Orlov--Schulman functions, and show that the factorization problem
leads to string equations. Thus all solutions of the Whitham
hierarchy fulfill certain set of string equations. In \cite{futuro}
we show that any solution of the string equations is a solution of
the Whitham hierarchy. We finish this section by introducing some
particular factorization problems and the corresponding string
equations which generalize and contains as particular cases, the
string equations of Krichever and of Takasaki--Takebe. Finally, \S 5
is devoted to the study of additional symmetries of the Whitham
hierarchy. First we derive the additional symmetries from the
factorization problem, and then characterize its action over the
potential function of the hierarchy. We compute the additional
symmetries of the mentioned Boyer--Finley--Benney system and obtain
a set of explicit functional symmetries. In particular, for the
Zakharov extension of the Benney system we get explicit functional
symmetries depending on three arbitrary functions of variable. We
conclude by considering the action of Virasoro type of additional
symmetries on our extension of the Krichever and Takasaki--Takebe
string equations and showing that solutions of string equations are
invariant solutions under a Lie algebra of additional symmetries,
which contains two set of Virasoro algebras.

\section{The factorization problem}

\subsection{Lie algebraic setting}
We  present a splitting  which is inspired in \cite{semenov} and in
\cite{manas1}, where it was used for a better understanding of
harmonic maps and chiral models. The factorization problem technique
was applied to the dispersionless KP hierarchy in \cite{39}, and is
inspired in the dressing method proposed by Takasi and Takabe in the
series of papers \cite{24}-\cite{21}.

Given the set $\{q_\mu^{(0)}\}_{\mu=0}^M\subset\bar\C$,
$q_0^{(0)}=\infty$, of punctures in the extended complex plane, we
introduce the local parameters $p_\mu^{-1}$ where
\[
p_\mu=\begin{cases} p, &\mu=0,\\(p-q_i^{(0)})^{-1}, & \mu=i\in\s,
\end{cases}
\]
and
\[
\s:=\{1,\dots,M\}.
\]
For each set of punctures we consider the set $\mathcal R$ of
rational functions in $p$ with poles at the punctures; i.e., the
functions $f=f(p)$ of the form
\[
f:=\sum_{\mu=0}^M\sum_{n=0}^{N_\mu}a_n^\mu p_\mu^{n},
\]
where $N_\mu\in\mathbb{N}$. In this paper we use  Greek letters like
$\mu$ to denote an index that runs from $0$ to $M$, and Italic
letters like $i$, when it runs from $1$ to $M$.

For each puncture $q_\mu^{(0)}$ we consider the set $\mathcal
L_\mu=\C(p_\mu)$ of formal Laurent series in $p_\mu$ and the subset
$\mathcal L^-_\mu$ defined as
\[
\mathcal L^-_\mu:=\begin{cases}  p^{-1}\C[[p^{-1}]],& \text{ for $\mu=0$,}\\
\C[[p-q_i^{(0)}]],& \text{ for  $\mu=i\in\s$}.
\end{cases}
\]
Here $\C[[p]]$ denotes the set of formal power series in $p$.
Finally, we define
\[
\mathcal L:=\bigoplus_{\mu=0}^M\mathcal L_\mu,\quad \mathcal
L^-:=\bigoplus_{\mu=0}^M\mathcal L_\mu^-.
\]

Given an element $(f_0,f_1,\dots,f_M)\in\mathcal L$,  let
$f_{(\mu,+)}$ be the polynomial in $p_\mu^{-1}$ such that $\tilde
 f_\mu^-:=f_\mu-f_{(\mu,+)}\in \mathcal L_\mu^-$. Then,  there exists a
unique rational function $f\in\mathcal R$ whose principal parts at
$q_\mu^{(0)}$ are given by $f_{(\mu,+)}$ (observe the normalization
condition at $\infty$), namely
\[
f=\sum_{\mu=0}^M f_{(\mu,+)}.
\]

Moreover, we have a unique splitting of $f_\mu$ of the form
\[
f_\mu=f_\mu^-+f
\]
with
\[
f_\mu^-:=\tilde f_\mu^--\sum_{\nu\neq\mu}\tilde f_{\nu}^-\in
\mathcal L_\mu^-.
\]
Therefore,  we conclude that the following splitting
\begin{equation}\label{split}
\mathcal L=\mathcal L^-\oplus \mathcal R
\end{equation}
holds.

\begin{figure}
\begin{center}
\psfrag{c01}{$q_1^{(0)}$}\psfrag{c02}{$q_M^{(0)}$}
\psfrag{c1}{$q_1$}\psfrag{c2}{$q_M$}\psfrag{c3}{$\infty$}
\psfrag{g1}{$\gamma_{1}$}\psfrag{g2}{$\gamma_{M}$}\psfrag{g3}{$\gamma_0$}
\psfrag{U1}{$D_{1}$}\psfrag{U2}{$D_{M}$}\psfrag{U3}{$D_0$}
\psfrag{L1}{$\mathcal L^-_{1}$}\psfrag{L2}{$\mathcal
L^-_{2}$}\psfrag{L3}{$\mathcal L^-_0$} \psfrag{C}{$\bar \C$}
\psfrag{R}{$\mathcal R$}
\includegraphics[height=10cm]{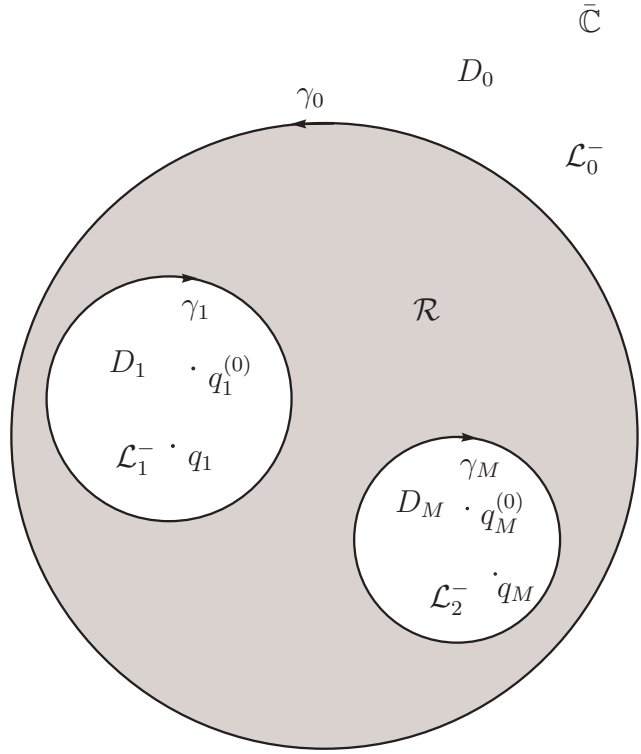}
\end{center}\caption{Graphical illustration of the splitting.}\label{diagram}
\end{figure}

The above construction can be extended in the following manner. Let
us consider, for $i\in\s$, the disk $D_i$ containing the point
$q_i^{(0)}$  with border the clockwise oriented circle
$\gamma_i:=\partial D_i$, and also the disk $D_0$, centered at $0$,
which contains all the other disks $D_i$, $i=1,\dots,M$,  with
border the counter-clockwise oriented circle $\gamma_0:=\partial
D_0$. Let $D:=D_0^c\cup\big(\bigcup_{i=1}^MD_i\big)$ and
$\gamma:=\bigcup_{\mu=0}^M\gamma_\mu$, here $D_0^c:=\bar
\C\backslash D_0$ is the complementary set of the disk $D_0$. We
will consider the completion of $\mathcal L$ as the set of complex
functions over $\Gamma$. We complete the rational splitting by
extending $\mathcal L_\mu^-$  as those complex functions over
$\gamma_\mu$ which admit analytic extensions to its interior,  and
for $\mu=0$ such that the extension vanishes at $\infty$. Then,
$\mathcal L^-=\oplus_{\mu=0}^M\mathcal L_\mu^-$ and $\mathcal R$ is
the set of complex functions on $\Gamma$ such that they do have an
holomorphic extension to $\C\setminus D$.
 In this context \eqref{split} also holds.
We refer to Figure \ref{diagram} for a graphical illustration of the
rational splitting and its completion.
Now, we shall extend the above splitting to the Lie algebra of
symplectic vector fields. In spite that normally the coordinates
$(p,x)$ are real, here we will consider that they take complex
values. This extension does not affect the standard local symplectic
constructions.

 The local Hamiltonian vector fields
\[ \boldsymbol{X}=A(p,x)\frac{\partial}{\partial
p}+B(p,x)\frac{\partial}{\partial x},
\]
are the divergence free vector fields $ A_p+B_x=0$, and locally
there exists a Hamiltonian function $H$ such that
\[
A=-\frac{\partial H}{\partial x},\quad B=\frac{\partial H}{\partial
p}.
\]
The Poisson bracket in the set $\mathcal F$ of differentiable
functions of $p$ and $x$  
is locally given by
\[
\{H,\tilde H\}=\frac{\partial H}{\partial p}\frac{\partial \tilde
H}{\partial x}-\frac{\partial H}{\partial x}\frac{\partial \tilde
H}{\partial p},
\]
and the pair $\g:=(\mathcal F,\{\cdot,\cdot\})$ is a Lie algebra.
The set of inner derivations of $\g$
\[
\ad H:=\{H,\cdot\}=\frac{\partial H}{\partial p}\frac{\partial
}{\partial x}-\frac{\partial H}{\partial x}\frac{\partial }{\partial
p}=\boldsymbol{X}_H
\]
may be locally identified with the set of Hamiltonian vector fields.
If fact, the set of locally Hamiltonian vector fields constitute a
Lie algebra under the Lie bracket given by the Lie derivative of
vector fields, and we have that
\[
[\boldsymbol{X}_H,\boldsymbol{X}_{\tilde
H}]=\boldsymbol{X}_{\{H,\tilde H\}},
\]
so that the mapping $H\to\boldsymbol{X}_H$ is a Lie algebra
homomorphism with kernel given by the constant functions; i.e., the
center of the Lie algebra of Hamiltonian functions.

We denote by $\mathfrak \g_\mu,\mathfrak \g_\mu^{-}$,
 and $\mathfrak r$,  the Lie subalgebras of $\g$
such that the corresponding Lie algebras of Hamiltonian vector
fields $\ad\mathfrak \g_\mu,\ad\mathfrak \g_\mu^{-}$ and
$\ad\mathfrak r$ are built up from vector fields with coefficients
in $\mathcal L_\mu$, $\mathcal L^-_\mu$ and $\mathcal R$,
respectively. Here, we suppose that the coefficients of the vector
fields are complex valued functions. Let us describe in more detail
these Lie algebras:
  \begin{enumerate}
\item \emph{The Lie algebra $\mathfrak r$ }. The components $A$ and $B$ of a  Hamiltonian vector field
\[
\ad H=A\frac{\partial}{\partial p}+B\frac{\partial}{\partial
x}\in\ad \r
 \]
are $A=-H_x$ and $B=H_p$ with
\[
H=\sum_{n=0}^{N_0}h_n(x)p^n+\sum_{i=1}^M\Big(h_{i0}
\log(p-q_i^{(0)})+\sum_{j=1}^{N_i}\frac{h_{ij}(x)}{(p-q_i^{(0)})^{j}}\Big),
\]
and $h_{i0,x}=0$.

\item \emph{The Lie algebras $\g_\mu$}.
The components $A_\mu$ and $B_\mu$ of a Hamiltonian vector field
\[
\ad H_\mu=A_\mu\frac{\partial}{\partial
p}+B_\mu\frac{\partial}{\partial x}\in\ad \g_\mu
 \]
are $A_\mu=-H_{\mu,x}$ and $B_\mu=H_{\mu,p}$ with
\[
 H_\mu=h_{\mu 0}\log(p_\mu)+\sum_{n\gg-\infty}h_{\mu n}(x)p_\mu^{-n},\quad h_{\mu 0,x}=0.
\]

\item \emph{The Lie algebra $\g^-$ }. The components $A_\mu$ and $B_\nu$ of a vector field
\[
\ad H_\mu=A_\mu\frac{\partial}{\partial
p}+B_\mu\frac{\partial}{\partial x}\in\g_{\mu}^{-},
 \]
are $A_\mu=-H_{\mu,x}$ and $B_\mu=H_{\mu,p}$ with
\[
H_0=h_{00}\log p+\sum_{n=1}^{\infty}h_{0n}(x)p^{-n},\quad
H_i=\sum_{n=0}^{\infty}h_{in}(x)(p-q_i^{(0)})^n,
\]
with $h_{00,x}=0$.
\end{enumerate}

Now, we define the Lie algebras
\[
 \g:=\bigoplus_{\mu=0}^M\g_\mu,\quad \g^-:=\bigoplus_{\mu=0}^M\g^-_\mu,
  \]
and realize that, modulo constants, the splitting \eqref{split}  in
this context is
\[
\ad\g=\ad\g^-\dotplus\ad\r
\]
which in turn is equivalent to
\[
\g=\g^-\dotplus\r.
\]

 The Lie algebras $\g_i$, for $i=1,\dots,M$, have a further splitting
into three Lie subalgebras:
\[
\g_i^-=\g_i^0\dotplus \g_i^1\dotplus \g_i^>,
\]
where
\[
 \quad\g_i^0:=\{h_{i0}(x)\},\quad\g_i^1=\{ h_{i1}(x)(p-q^{(0)}_i)\},\quad \g_i^>:=\{h_{i2}(x)
 (p-q^{(0)}_i)^2+h_{i3}(x)(p-q_i^{(0)})^3+\cdots\}
\]
and $\{\g_i^0\dotplus\g_i^1,\g_i^>\}\subset\g_i^>$. The above
splitting induces the following splitting into Lie subalgebras of
divergence free vector fields
\[
\ad\g_i^-=\ad\g_{i}^0\dotplus \ad\g_{i}^1\dotplus\ad\g_{i}^{>}.
\]

\subsection{Lie group setting}

We now extend the previous construction from the context of Lie
algebras to the corresponding Lie groups of canonical
transformations. Associated with each Hamiltonian vector field
$\boldsymbol X_H=\ad H$ we have the corresponding Hamilton's
equations $\dot p=-H_x,\dot x=H_p$ that when integrated provides us
with a flow $\Phi_t^{H}$, a 1-parameter group of symplectic
diffeomorphims, $(p(t),x(t))=\Phi_t^{H}(p_0,x_0)$ for given initial
conditions $(p,x)|_{t=0}=(p_0,x_0)$. The exponential mapping is just
the evaluation at $t=1$; i.e. $\exp\boldsymbol X_H=\Phi^H_{t=1}$.
 The group of symplectic diffeomorphims is a smooth regular Lie group with Lie
algebra given by the set of Hamiltonian vector fields \cite{michor}.
Symplectic diffeormorphims are also known as canonical
transformations.

It can be shown \cite{michor} that the adjoint action of the group
of symplectic diffeomorphims on its Lie algebra (i.e., the set of
Hamiltonian vector fields) is given by the action of the
corresponding induced flow:
\begin{gather}\label{Adjoint}
\begin{aligned}
\Ad\exp(s\boldsymbol X_H)(\boldsymbol X_{\tilde
H})&=(\Phi_{-s}^H)^*\boldsymbol X_{\tilde
H}=T\Phi_{s}^H\circ\boldsymbol X_{\tilde
H}\circ\Phi_{-s}^H=
\boldsymbol X_{(\Phi _{-s}^H)^*\tilde H}=\boldsymbol
X_{\Ad\exp(sH)\tilde H},
\end{aligned}
\end{gather}
where
\[
\Ad\exp(sH)\tilde H:=(\Phi _{-s}^H)^*\tilde H=\Exp{s\ad H}\tilde
H=\sum_{l=0}^\infty\frac{(s\ad H)^l}{l!}\tilde H.
\]
That is, modulo constants, the adjoint action of a symplectic
diffeomorphism of the form $\exp(\boldsymbol X_H)$ acts on the
Hamiltonian functions as $\Exp{\ad H}$:
\[
\exp(\boldsymbol X_H)\xrightarrow{\Ad} \Exp{\ad H}.
\]

 The rational splitting of Lie algebras of Hamiltonian
vector fields may be exponentiated  to a Birkhoff type factorization
problem
\[
\exp({\boldsymbol X_\mu})=\exp(\boldsymbol
X_\mu^-)^{-1}\circ\exp({\boldsymbol X}) \text{ with $\boldsymbol
X_\mu\in\ad \g_\mu$, $\boldsymbol X_\mu^-\in\ad\g_\mu^{-}$ and
$\boldsymbol X\in\ad\r$},
\]
where  we are now dealing with complex vector fields.

We will consider a particular class of Hamiltonians, namely those of
the following form
\begin{gather}\label{tmu}
T_\mu:=(1-\delta_{\mu 0})t_{\mu 0}\log p_\mu+\sum_{n=1+\delta_{\mu
0}}^\infty t_{\mu n}p_\mu^{n}.
\end{gather}
Given  \emph{initial} canonical transformations $\Phi_\mu$,
$\mu=0,\dots, M$, we consider  deformations $\exp(\boldsymbol
X_{T_\mu})\circ\Phi_\mu$ which area new canonical transformations
that now depend on the deformation or \emph{time} parameters
\[
\bt:=(t_{\mu n}).
\]
We will consider the factorization
\begin{gather}\label{factorization}
\exp({\boldsymbol{X}_{T_\mu}})\circ\Phi_\mu=\exp({\boldsymbol
X_\mu^-})^{-1}\circ\exp({\boldsymbol X}) \text{ with  $\boldsymbol
X_\mu^-\in\ad\g_\mu^{-}$ and $\boldsymbol X\in\ad\r$}.\end{gather}
Equation \eqref{factorization} is fulfilled if the following
factorization problem is satisfied
\begin{gather}\label{factorization2}
\Exp{\ad T_\mu}\Exp{\ad G_\mu}=\Exp{-\ad H_\mu^-}\Exp{\ad
H},\text{ with  $H_\mu^-\in\g_\mu^{-}$ and $H\in\r$},
\end{gather}
where,
\[
\Phi_\mu =\exp{\boldsymbol{X}_{G_\mu}},\; \boldsymbol{
X}_\mu^-=\boldsymbol{X}_{H_\mu^-},\;
\boldsymbol{X}=\boldsymbol{X}_{H}.
\]

 The existence
problem for \eqref{factorization2} will not be treated here.
Anyhow, we will assume that all times $|t_{\mu n}|$ and initial
conditions are small enough to ensure that such factorization
exists (notice the trivial existence for $T_\mu=0$ and $G_\mu=0$).

Observe that given a set of \emph{initial conditions}
$\{G_\mu\}_{\mu=0}^M$ the factorization problem
\eqref{factorization2} consists in finding $H_\mu^-$,
$\mu=0,\dots,M$, and $H$ as functions of $\bt$. Let us
right-multiply both terms of the equality \eqref{factorization2} by
a term of the form $\Exp{\ad G}$, where $G\in\r$. On the left-hand
term we have $\Exp{\ad T_\mu}\Exp{\ad \tilde G_\mu}$, where the new
initial conditions are
\begin{gather}\label{campbell}
\tilde
G_\mu:=C(G_\mu,G):=G_\mu+G+\frac{1}{2}\{G_\mu,G\}+\frac{1}{12}(\{G_\mu,\{G_\mu,G\}+\{G
,\{G,G_\mu\}\})+\cdots,
\end{gather}
and $C(\cdot,\cdot)$ is the Campbell--Hausdorff series in Dynkin
form, so that
\begin{gather}\label{invariance}
\Exp{\ad G_\mu}\Exp{\ad G}=\Exp{\ad\tilde G_\mu}.
\end{gather}
 A solution of
this new factorization problem is given by $\tilde H^-_\mu=H^-_\mu$
and $\tilde H=C(H,G)\in\r$, so that $H_\mu^-$ remains invariant. Let
us now left-multiply both terms of the equality by $\Exp{\ad
c^-_\mu(p)}$, with $c^-_\mu\in\mathfrak c^-_\mu$, being $\mathfrak
c_\mu^-\subset \g_\mu^-$  the Abelian subalgebra of Hamiltonians in
$\g_\mu^-$ which only depend on $p$. As $\{c_\mu^-,T_\mu\}=0$ we
have $\Exp{\ad c_\mu^-}\Exp{\ad T_\mu}\Exp{\ad G_\mu}=\Exp{\ad
T_\mu}\Exp{\ad \tilde G_\mu}$ with $\tilde G_\mu:=C(c_\mu^-,G_\mu)$.
The solution of the transformed factorization problem
\eqref{factorization2} is given by $\tilde
H_\mu^-=C(c_\mu^-,H^-_\mu)$ and $\tilde H=H$.

Therefore, once we have a solution $(H_\mu^-,H)$ for an initial
condition $G_\mu$ it is trivial to find solutions $(\tilde
H_\mu^-,\tilde H)$ for initial conditions $C(\mathfrak
c_\mu^-,C(G_\mu,\r))$. The orbits $\Exp{\ad\mathfrak
c_\mu^-}\Exp{\ad G_\mu}\Exp{\ad\r}$ describe the moduli space of
solutions to the factorization problem \eqref{factorization2}. Thus,
if we concentrate on the right action of $\r$, we may take
$G_\mu\in\g_\mu^-$ and the right coset $\Exp{\ad G_\mu}\Exp{\ad\r}$
(or the Hamiltonian $C(G_\mu,\r)$) as the point in the moduli.

As we will see the factorization problem \eqref{factorization2} for
the action of symplectic diffeomorphims on the set of functions
(observables) implies the Whitham hierarchy. Therefore, the
factorization problem \eqref{factorization} for symplectic
diffeormorphism  is also associated with the Whitham hierarchy. To
get these results we will use a well known tool in the theory of
regular Lie groups:  the right logarithmic derivative as defined in
\cite{michor}, see the Appendix. If we have a smooth curve
$H:\mathcal{T}\rightarrow C^\infty(\mathcal N)$,  assuming that
$\mathcal{T}$ is the time
 manifold with local coordinates given by $\bt=(t_{\mu n})$ and
 denoting $\partial_{\mu m}:=\pde{}{t_{\mu n}}$, the right logarithmic derivative
is
\[
\delta\exp(\boldsymbol X_H)(\partial_{\mu
n})=\int_0^1(\Phi_{-s}^H)^*(T_{\bt}\boldsymbol X_H(\partial_{\mu
n}))\d s=\boldsymbol X_{\delta \Exp{\ad H}(\partial_{\mu n})},
\]
where
\[
\delta \Exp{\ad H}(\partial_{\mu
n}):=\int_0^1(\Phi_{-s}^H)^*(\partial_{\mu n}H)\d
s=\sum_{l=0}^\infty\frac{(\ad H)^l}{(l+1)!}\partial_{\mu n}\tilde H.
\]
In particular,
\[
\delta\exp({\boldsymbol X_{T_\mu}})(\partial_{\mu n})=\boldsymbol
X_{\frac{\partial T_\mu(p)}{\partial t_{\mu n}}} \text{ with }
\frac{\partial T_\mu(p)}{\partial
t_{\mu n}}=\begin{cases}   p_\mu^{n},& n\neq 0\\
\log(p-q_i^{(0)}), & n=0, \mu =i.
\end{cases}
\]
Now, we are ready to take right logarithmic derivatives, using
\eqref{rules}, of the factorization problem \eqref{factorization},
\begin{gather}\label{rd-1}
\begin{aligned}
\delta\exp(\boldsymbol X_\nu^-)(\partial_{\mu
n})+\delta_{\mu\nu}\Ad\exp(\boldsymbol X_\mu^-)\big(\boldsymbol
X_{\frac{\partial T_\mu(p)}{\partial t_{\mu
n}}}\big)&=\delta\exp({\boldsymbol
X})(\partial_{\mu n}),
\end{aligned}
\end{gather}
Which, using the corresponding Hamiltonian generators
\[ \boldsymbol
X_\mu^-=\boldsymbol X_{H_\mu^-},\quad \boldsymbol X=\boldsymbol
X_{H}
\]
we get, modulo constants, the following system
 \begin{equation}
   \label{rd}
   \begin{aligned}
   \delta\Exp{\ad H_\nu^-}(\partial_{\mu n})+\delta_{\mu\nu}\Exp{\ad H_\mu ^-}\Big(\dfrac{\partial T_\mu(p)}{\partial t_{\mu n}}\Big)&=
   \delta\Exp{H}(\partial_{\mu n}),
   \end{aligned}
 \end{equation}
which may be derived directly from \eqref{factorization2} by taking
right logarithmic derivatives.

\section{Dressing methods for the Whitham hierarchy}

In this section we analyze how the factorization problem
\eqref{factorization2} is related with the Whitham hierarchy and its
dressing transformations. We first show that \eqref{factorization2}
leads to the Whitham hierarchy, defining the Lax functions a
zero-curvature forms. Then, we construct a potential function
$h_{01}$ of this hierarchy, and as we shall show in the forthcoming
paper \cite{futuro} $h_{01}=-(\log\tau)_x$ in terms of the
$\tau$-function of the hierarchy.  We also proof that any solution
of the Whitham hierarchy is related to a factorization problem, via
an undressing procedure. Finally, we show how the factorization
problem scheme can be extended to generate dressing transformations
of the Whitham hierarchy.

\subsection{From the factorization problem to the Whitham
hierarchy}\label{dressing}

We are now ready to proof  that \eqref{factorization2} is described
differentially by the Whitham hierarchy
\begin{teh}\label{Theorem-fac-whitham}
  Given a solution of the factorization problem
  \eqref{factorization2},
  \[
\Exp{\ad T_\mu}\Exp{\ad G_\mu}=\Exp{-\ad H_\mu^-}\Exp{\ad H},\quad
H_\mu^-\in\g_\mu^-,\; H\in\r,
  \]
  then:
  \begin{enumerate}
    \item The Lax functions
    \begin{gather}\label{laxf}
    z_\mu:=\Exp{\ad H_\mu^-}p_\mu
    \end{gather}
     are of the
    form
    \begin{gather}\label{lax_functions}
      z_\mu=\begin{cases}
        p+\sum_{l=1}^\infty d_{0l}p^{-l}, & \mu=0,\\
        \dfrac{d_{i-1}}{p-q_i}+\sum_{l=0}^\infty d_{il} (p-q_i)^l,&
        \mu=i\in\s.
      \end{cases}
    \end{gather}
    for some functions $q_i$ and $d_{\mu m}$ defined in terms of the
    coefficients of $H_\mu^-$.
    \item The functions
    \begin{gather}\label{def-omega}
\Omega_{\mu n}:=\begin{cases}   (z_\mu^n)_{(\mu,+)} ,& n>\delta_{\mu
0},\\
-\log(p-q_i), &n=0,\quad \mu=i\in\s
\end{cases}
    \end{gather}
    where $(\cdot)_{(i,+)}$ projects in the span $\{\log(p-
q_i),(p- q_i)^{-l}\}_{l=1}^\infty$ and $(\cdot)_{(0,+)}$ onto the
span of $\{p^l\}_{l=0}^\infty$, satisfy the zero-curvature equations
\begin{gather}
  \label{whitham}\pde{\Omega_{\mu n}}{t_{\nu l}}-\pde{\Omega_{\nu l}}{t_{\mu
n}}+\{\Omega_{\mu n},\Omega_{\nu l}\}=0,
\end{gather}
moreover
\[
\Omega_{\mu n}=\delta\Exp{\ad H}(\partial_{\mu n}).
\]
\item The Lax functions $z_\mu$ are subject to the Whitham hierarchy:
\begin{gather}
  \label{lax-whitham}\pde{z_\nu}{t_{\mu n}}=\{\Omega_{\mu n},z_\nu\}.
\end{gather}

  \end{enumerate}

\end{teh}
\begin{proof}
 We now proceed to show that \eqref{rd} implies the Whitham
hierarchy. In the analysis of \eqref{rd} is convenient to
distinguish between the cases $\mu=i\neq 0$ and $\mu=0$.

\begin{enumerate}
\item \textbf{The case  $\mu=i\in\s$.}

We factor $\Exp{\ad H_i^-}=\Exp{\ad H_{i0}}\Exp{\ad H_{i1}}\Exp{\ad
H_{i>}}$, with
\[
H_{i0}=h_{i0}(x),\quad H_{i1}=h_{i1}(x)(p-q_i^{(0)}),\quad
H_{i>}=h_{i2}(x)(p-q_i^{(0)})^2+h_{i3}(x)(p-q_i^{(0)})^3+\cdots.
\]
Now, we study the cases $m>0$ and $m=0$:

\begin{enumerate}
\item
\textbf{$m>0$ }

  We get
\begin{gather} \label{rd2}
  \delta \Exp{\ad H_{i0}}(\partial_{in})+\Exp{\ad H_{i0}}
  \Big( \delta \Exp{\ad H_{i1}}(\partial_{in})+
\Exp{\ad H_{i1}}\big(\delta \Exp{\ad
H_{i>}}(\partial_{in})\big)\Big)+z_i^n
 =\delta \Exp{\ad H}(\partial_{in})
 \end{gather}
It can be proved that
\begin{align*}
   \delta \Exp{\ad H_{i0}}(\partial_{in})&=\partial_{in}h_{i0},\\
  \delta \Exp{\ad H_{i1}}(\partial_{in})&=\frac{\partial_{in}X_i}{X_{i,x}}(p-X_{i,x}q_i^{(0)}),\quad
   \text{with $\int_x^{X_i}\frac{\d
   x}{h_{i1}(x)}=1$},\\
\delta \Exp{\ad
H_{i>}}(\partial_{in})&=\partial_{in}h_{i2}(p-q_i^{(0)})^2+(\partial_{in}h_{i3}
+h_{i2}\partial_{in}h_{i2,x}-h_{i2,x}\partial_{in}h_{i2})(p-q_i^{(0)})^3+\cdots,\\
    \Exp{\ad H_{i0}}(f(x)(p-q_i^{(0)})^n)&=f(x) (p-q_i^{(0)}-h_{i0,x})^n,\\
    \Exp{\ad H_{i1}}(f(x)(p-q_i^{(0)})^n)&=\frac{f|_{x=X_i}}{(X_{i,x})^n}\big(p-X_{i,x}q_i^{(0)}\big)^n,\text{ and in particular
    $\Exp{\ad H_{i1}}x=X_i$}\\
   \Exp{\ad H_{i>}}\Big(\frac{1}{(p-q_i^{(0)})^n}\Big)&=\Big(\frac{1}{p-q_i^{(0)}}+h_{i2,x}
+(h_{i3,x} +h_{i2}h_{i2,xx})(p-q_i^{(0)})+\cdots\Big)^n.
\end{align*}
Therefore, defining
\[
    q_i:=X_{i,x}q_i^{(0)}+h_{i0,x},
\]
we deduce that \eqref{rd2} can be written as
\begin{gather}
   \label{rd3}
  \partial_{in}h_{i0}+\frac{\partial_{in}X_i}{X_{i,x}}(p- q_i)
  +\frac{(\partial_{in}h_{i2})|_{x=X_i}}{(X_{i,x})^2}(p- q_i)^2+\cdots+
  z_i^n=\delta \Exp{\ad H}(\partial_{in})
 \end{gather}
with
\[
z_i:=\Exp{\ad
 H_i^-}\Big(\frac{1}{p-q_i^{(0)}}\Big)=
 \frac{d_{i\,-1}}{p- q_i}+\sum_{l=0}^\infty d_{il} (p-q_i)^l,
\]
where, for example,
\[
d_{i\,-1}:=X_{i,x},\quad d_{i0}:=h_{i2,x}|_{x=X_i},\quad
d_{i1}:=\frac{(h_{i3,x} +h_{i2}h_{i2,xx})\big|_{x=X_i}}{X_{i,x}}.
\]

We have assumed  that $T_\mu$ and $G_\mu$ are small enough to
ensure that  the function $ q_i=X_{i,x}q_i^{(0)}+h_{i0,x}$ belongs
to the interior of $U_{q_i^{(0)}}$ (so that $X_{i,x}\approx 1$ and
$h_{i0,x}\approx 0$). Thus, \eqref{rd3} implies
\[
\r\ni\delta\Exp{\ad
H}(\partial_{im})=(z_i^m)_{(i,+)}=:\Omega_{im}.
\]

For example,
\[
\Omega_{i1}=\frac{d_{i\,-1}}{p- q_i},\quad
\Omega_{i2}=\frac{d_{i\,-1}^2}{(p-
q_i)^2}+\frac{2d_{i\,-1}d_{i0}}{p- q_i}.
\]
\item\textbf{$m=0$}

In this case we have
\begin{gather}
   \label{rd4}
  \partial_{i0}h_{i0}+\frac{\partial_{i0}X_i}{X_{i,x}}(p- q_i)
  +\frac{(\partial_{i0}h_{i2})|_{x=X_i}}{(X_{i,x})^2}(p- q_i)^2+\cdots+
  \log z_i=\delta \Exp{\ad H}(\partial_{i0}).
 \end{gather}
 Notice that
\begin{gather*}
\log z_i:=-\log(p- q_i)+\log\Big(X_{i,x}+h_{i2,x}|_{x=X_i}(p- q_i)
+\frac{(h_{i3,x} +h_{i2}h_{i2,x})|_{x=X_i}}{X_{i,x}}(p-
q_i)^2+\cdots\Big),
\end{gather*}
and hence
\[
\r\ni\delta \Exp{\ad H}(\partial_{i,0})=(\log
z_i)_{(i,+)}=:\Omega_{i0}
\]
with
\[
\Omega_{i0}=-\log(p- q_i).
\]

\end{enumerate}
\item \textbf{$\mu=0$ }

In this case we have
\begin{equation}
   \label{rdinfty}
  \delta\Exp{\ad H_0^-}(\partial_{0n})+z_0^n=
   \delta\Exp{\ad H}(\partial_{0n}).
 \end{equation}
 with
 \[
 \Exp{\ad H_0^-}=\Exp{\ad H_{0>}}\Exp{\ad(t_{00}\log p)},\quad H_{0>}=h_{01}
 p^{-1}+h_{02} p^{-2}+\cdots,
 \]
 where $t_{00}$, which is not a time parameter,
does not depend on $x$. Notice that
\[
z_0=\Exp{\ad H_0^-}(p)=p+\sum_{l=1}^\infty d_{0l}p^{-l}
\]
where, for example,
\[
d_{01}:=-h_{01,x},\quad d_{02}:=-h_{02,x}. \]

An analysis of equation \eqref{rdinfty} allows us to write
\[
\r\ni\delta\Exp{\ad H}(\partial_{0n})=(z_0^n)_{(0,+)}=:\Omega_{0n},
\]
for example
\[
\Omega_{02}=p^2+2d_{01}.
\]

From
\[
\Omega_{\mu n}=\delta\Exp{\ad H}(\partial_{\mu n})
\]
and \eqref{zc} we deduce the zero-curvature conditions
\eqref{whitham}.
\item From \eqref{dz} we have
\[
\pde{z_\nu}{t_{\mu n}}=\{\delta\Exp{\ad H_\nu^-}(\partial_{\mu
n}),z_\nu\}
\]
that recalling \eqref{rd} reads
\[
\pde{z_\nu}{t_{\mu n}}=\{\delta\Exp{\ad H}(\partial_{\mu n}),z_\nu\}
\]
and we deduce \eqref{lax-whitham}.
\end{enumerate}\end{proof}
As a byproduct of the above proof we have the following
\begin{pro}\label{fac-coefficients}
Given solutions $H_\mu^-$ and $H$ of the factorization
  problem \eqref{factorization2} such that
\begin{align*}
\Exp{\ad H_\mu^-}=\begin{cases} \Exp{\ad (\sum_{l=1}^\infty
h_{0l}(x)p^{-l})}\Exp{\ad(t_{00}\log p)}, &\mu=0,\\
\Exp{\ad h_{i0}(x)}\Exp{\ad h_{i1}(x)(p-q_i^{(0)})}\Exp{\ad
(\sum_{l=2}^\infty h_{il}(x)(p-q_i^{(0)})^l)},& \mu=i\in\s,
\end{cases}
\end{align*}
then \begin{gather}\label{t00} t_{00}=-\sum_{i=1}^Mt_{i0},
\end{gather}
 and the coefficients of the Lax
functions satisfy
\begin{align*}
q_i&=X_{i,x}q_i^{(0)}+h_{i0,x},\quad  \int_x^{X_i}
\dfrac{\d x}{h_{i1}(x)}=1,& &\\
d_{i\,-1}&=X_{i,x},&&\\
d_{il}&=(h_{i\,l+2,x}+f_{il}(h_{i\,l+1},\dots,h_{i2}))|_{x=X_i}X_{i,x}^{-l},& l&\geq 0\\
d_{0l}&=-h_{0l,x}+f_{0l}(h_{0\,l-1},\dots,h_{01}),&&\\
\end{align*}
where $f_{\mu l}$ are differential polynomials.
\end{pro}

\begin{proof}
  We only need to prove \eqref{t00}. We will consider the equations
  \begin{gather}
    \label{t000n}
    \begin{aligned}
      \delta\Exp{\ad H_{0>}}(\partial_{\mu n})+\pde{t_{00}}{t_{\mu n}}\log
      z_0+\delta_{\mu 0 }z_0^n&=\delta\Exp{\ad H}(\partial_{\mu n}),\\
 \delta\Exp{\ad H_i^-}(\partial_{\mu n})+\delta_{\mu i}((1-\delta_{n0})z_i^n+ \delta_{n0}\log z_i)&=\delta\Exp{\ad
 H}(\partial_{\mu n}),\quad i\in\s,
    \end{aligned}
  \end{gather}
    which are derived from \eqref{factorization2} by taking right
  logarithmic derivatives. 

  We take the $p$-derivative of \eqref{t000n} to get
 \begin{gather}
    \label{t000nd}
    \begin{aligned}
      \frac{\d}{\d p}\Big[\delta\Exp{\ad H_{0>}}(\partial_{\mu n})\Big]
      +\Big(\pde{t_{00}}{t_{0n}}\frac{1}{z_0}+n\delta_{\mu 0}z_0^{n-1}\Big)\frac{\d z_0}{\d p}&=
      \frac{\d}{\d p}\Big[\delta\Exp{\ad H}(\partial_{\mu n})\Big],\\
 \frac{\d}{\d p}\Big[\delta\Exp{\ad H_i^-}(\partial_{\mu n})\Big]+\delta_{\mu i}\Big(n(1-\delta_{n0})z_i^{n-1}+
 \delta_{n0}\frac{1}{z_i}\Big)\frac{\d z_i}{\d p}&=\frac{\d}{\d p}\Big[\delta\Exp{\ad
 H}(\partial_{\mu n})\Big],\quad i\in\s.
    \end{aligned}
  \end{gather}
Now,
\[
 \frac{\d}{\d p}\Big[\delta\Exp{\ad
 H}(\partial_{\mu n})\Big]
\]
is analytic in $\bar\C\backslash D$ and therefore
\[
0=\oint_\gamma\frac{\d}{\d p}\Big[\delta\Exp{\ad
 H}(\partial_{\mu n})\Big]\d p=\sum_{\mu=0}^M\oint_{\gamma_\mu}\frac{\d}{\d p}\Big[\delta\Exp{\ad
 H}(\partial_{\mu n})\Big]\d p
\]
but from \eqref{t000nd} we deduce
\begin{align*}
  \oint_{\gamma_0}\frac{\d}{\d p}\Big[\delta\Exp{\ad
 H}(\partial_{\mu n})\Big]\d p&=  \oint_{\gamma_0}\frac{\d}{\d p}\Big[\delta\Exp{\ad
 H_{0>}}(\partial_{\mu n})\Big]\d p
      + \oint_{\Gamma_0}\Big(\pde{t_{00}}{t_{\mu n}}\frac{1}{z_0}+n\delta_{\mu 0}z_0^{n-1}\Big)\d z_0,\\
\oint_{\gamma_i}\frac{\d}{\d p}\Big[\delta\Exp{\ad
 H}(\partial_{\mu n})\Big]\d p&=\oint_{\gamma_i}\frac{\d}{\d p}\Big[\delta\Exp{\ad
 H_i^-}(\partial_{\mu n})\Big]\d p+\delta_{\mu i}\oint_{\Gamma_i} \Big(n(1-\delta_{n0})z_i^{n-1}+
  \delta_{n0}\frac{1}{z_i}\Big)\d z_i,\quad i\in\s,
\end{align*}
where we have changed of variables $z_\mu=z_\mu(p)$ with
$\Gamma_\mu=z_\mu(\gamma_\mu)$. Now, recalling that
\begin{gather*}
 \frac{\d}{\d p}\Big[\delta\Exp{\ad
 H_{0>}}(\partial_{\mu n})\Big]=O(p^{-2}),\quad p\to\infty,\\
\frac{\d}{\d p}\Big[\delta\Exp{\ad
 H_i^-}(\partial_{\mu n})\Big] \text{ is holomorphic at $D_i$ for $i\in\s$}
\end{gather*}
we get
\[
\pde{t_{00}}{t_{\mu n}}=-(1-\delta_{\mu 0})\delta_{n0}.
\]
\end{proof}


\subsection{Some dispersionless systems within the Whitham hierarchy: the Boyer--Finley--Benney
system}\label{Sistema}
  We consider the equations involving the times
  $\{t_{i0}=:x_i,t_{j1}=:y_j, t_{02}=:t\}_{i,j=\in\s}$. Now, we write
  \[
  \Omega_{i0}=-\log(p- q_i),\;\Omega_{i1}=\dfrac{v_i}{p-
  q_i}\text{ and } \Omega_{02}=p^2-2w, \text{ with } \quad
  v_i:=d_{i-1}, \text{ and }
  w:=-d_{01}.
  \]
and the corresponding Whitham equations \eqref{whitham}  are
\begin{align}
\frac{\partial  q_i}{\partial y_{j}}&=\frac{\partial v_j}{\partial
x_{j}}=\frac{\partial }{\partial x}\Big(\frac{v_j}{ q_i-
q_j}\Big),\label{bf1}\\
  \frac{\partial  q_i}{\partial y_{i}}&=\frac{\partial v_i}{\partial
x_{i}},\label{bf2}\\
\frac{\partial  q_i}{\partial x_{j}}&=-\frac{\partial
\log( q_i- q_j)}{\partial x}, \label{bf3}\\
\frac{\partial  q_i}{\partial x_{i}}&=-\frac{\partial
\log(v_i)}{\partial x},\label{bf4}\\
\pde{w}{x_{i}}&=\pde{q_i}{x},\label{bf7}\\
\pde{q_i}{t}&=\pde{(q_i^2-2w)}{x},\label{bf5}\\
\pde{v_i}{t}&=2\pde{(q_iv_i)}{x},\label{bf6}\\
\pde{w}{y_{i}}&=\pde{v_i}{x},\label{bf8}
\end{align}
where $i\neq j$.

Observe that equations \eqref{bf2} and \eqref{bf4} imply
\begin{gather}\label{bf}
 \frac{\partial^2 \Exp{\Phi_i}}{\partial x_{i}^2}+\frac{\partial^2
\Phi_i}{\partial x\partial y_{i}}=0, \quad \Phi_i:=\log v_i,
\end{gather}
which is the Boyer--Finley equation,  which appears in General
Relativity \cite{27}, or dispersionless Toda equation for $\Phi_i$,
and that equations \eqref{bf5}-\eqref{bf8} form the Benney
generalized gas system \cite{benney}.

Notice also that from \eqref{bf3}, \eqref{bf1}, \eqref{bf7} and
\eqref{bf8} we deduce the local existence of a potential function
$W$ such that
\[
 q_i=\frac{\partial W}{\partial x_{i}},\quad
 v_i=\frac{\partial W}{\partial y_{i}},\quad
 w=\pde{W}{x}.
\]
Therefore, this system of equations may be simplified as follows
\begin{align}
 W_{x_{i}y_{j}}-\Big(\frac{W_{y_{j}}}{W_{x_{i}}-W_{x_{j}}}\Big)_x&=0,&
i\neq j,\label{bfw1}\\
 W_{x_{i}x_{j}}+(\log(W_{x_{i}}-W_{x_{j}}))_x&=0, & i\neq j,\label{bfW2}\\
W_{x_{i}x_{i}}+(\log(W_{y_{i}}))_x&=0,&\label{bWf4}\\
W_{x_{i}t}+(2W_x-W_{x_{i}}^2)_x&=0,&\label{bWf5}\\
 W_{y_{i}t}-2(W_{x_{i}}W_{y_{i}})_x&=0,&\label{bWf6}
\end{align}

We stress again that  \eqref{bWf4} is a form of the Boyer--Finley
equation, and that \eqref{bWf5} and \eqref{bWf6} is a form of Benney
system. Therefore, the whole system may be understood as an
extension of these equations. This fact, have induce us to propose
the name of Boyer--Finley--Benney for the mentioned system.

\subsection{On the existence of a potential for the Whitham hierarchy}

In the previous section we have seen that the Boyer--Finley--Benney
equations can be reformulated in terms of a single field. We will
show now that this is a general fact for the Whitham hierarchy,
being the potential the coefficient
\[
 h_{01}=:-(\log\tau)_x,
 \]
  as we will see in a
forthcoming paper this is essentially due to the existence of a
$\tau$-function for the Whitham hierarchy \cite{futuro}.

The Whitham hierarchy is determined in terms of the functions
$z_\mu$ or its coefficients $d_{\mu n}$ as given in
\eqref{lax_functions}. In fact, as was stated in Proposition
\ref{fac-coefficients} the coefficients $d_{\mu n}$ are determined
in terms of $h_{\mu m}$ and its $x$-derivatives. We will consider
inversion formulae for \eqref{lax_functions}
  \begin{gather}\label{inversion}
     \begin{aligned}
      p&=z_0+\sigma_{01}z_0^{-1}+\sigma_{02}z_0^{-2}+\cdots,\\
      p&=q_i+\sigma_{i1}z_i^{-1}+\sigma_{i2}z_i^{-2}+\cdots,
     \end{aligned}
  \end{gather}
where the inversion coefficients $\sigma_{\mu n}$ are polynomials in
$d_{\mu m}$, for example
\begin{align}
\sigma_{01}&=-d_{01},& \sigma_{02}&=-d_{02},&
\sigma_{03}&=-(d_{03}+d_{01}^2),\\
\sigma_{i1}&=d_{i\,-1},& \sigma_{i2}&=d_{i0}d_{i\,-1},&
\sigma_{i3}&=d_{i\,-1}d_{i0}+d_{i\,-1}^2d_{i1}.
\end{align}



In the following we will use the geometry illustrated in Figure
\ref{diagram}. We first show the following

\begin{teh}\label{cauchy-pro}
The following identity holds
\begin{multline}
  \label{H0>}
\Big[\delta\Exp{\ad H_{0>}}(\partial_{\mu
n})\Big](p)=-\frac{1}{2\pi\I}\oint_{\Gamma_\mu}
\log\Big(1-\frac{p(z_\mu)}{p}\Big)nz_\mu^{n-1}\d
z_\mu\\+(1-\delta_{\mu 0})
\delta_{n0}\Big(\log\Big(1-\frac{q_\mu}{p}\Big)-\frac{1}{2\pi\I}\oint_{\Gamma_0}\log\Big(1-\frac{p(z_0)}{p}\Big)z_0^{-1}\d
z_0\Big),\quad p\in \bar\C\backslash D_0.
\end{multline}
\end{teh}
In the above formula we must understand that when $\mu=0$ the second
term of the r.h.s. vanishes even if $q_0=\infty$.
\begin{proof}
 We first introduce
  \[
\delta\Exp{\ad H_{0>}}(\partial_{\mu n})=:\Phi_{\mu n}=\Phi_{\mu
n,1}p^{-1}+\Phi_{\mu n,2}p^{-2}+\cdots
  \]
  and observe that
  \begin{gather}\label{phi}
\frac{1}{2\pi\I}\oint_{\gamma_0}p^m\frac{\d\Phi_{\mu n}}{\d p}(p)\d
p=-m\Phi_{\mu n,m},\quad m=1,2,\dots
  \end{gather}
Now we consider \eqref{t000n} with the explicit form for $t_{00}$
  \begin{gather}
    \label{t000n00}
    \begin{aligned}
      \delta\Exp{\ad H_{0>}}(\partial_{\mu n})+\delta_{\mu 0 }z_0^n-(1-\delta_{\mu 0}) \delta_{n0}\log
      z_0&=\delta\Exp{\ad H}(\partial_{\mu n}),\\
 \delta\Exp{\ad H_i^-}(\partial_{\mu n})+\delta_{\mu i}(
  (1-\delta_{n0})z_i^n+\delta_{n0}\log z_i)&=\delta\Exp{\ad
 H}(\partial_{\mu n}),\quad i\in\s
    \end{aligned}
  \end{gather}
  which are derived from \eqref{factorization2} by taking right
  logarithmic derivatives. We act with $p^m\dfrac{\d}{\d p}$ on
  \eqref{t000n00} to get
\begin{gather}
    \label{t000n001}
    \begin{aligned}
     p^m\dfrac{\d}{\d p}\Big[\delta\Exp{\ad H_{0>}}(\partial_{\mu n})\Big]+
     p^m(n\delta_{\mu 0 }z_0^{n-1}-(1-\delta_{\mu 0}) \delta_{n0}z_0^{-1})\frac{\d z_0}{\d p}&=p^m\dfrac{\d}{\d p}\Big[
     \delta\Exp{\ad H}(\partial_{\mu n})\Big],\\
 p^m\dfrac{\d}{\d p}\Big[\delta\Exp{\ad H_i^-}(\partial_{\mu n})\Big]
 +\delta_{\mu i}(n z_i^{n-1}+\delta_{n0}z_i^{-1}))\frac{\d z_i}{\d p}&=p^m\dfrac{\d}{\d p}\Big[\delta\Exp{\ad
 H}(\partial_{\mu n})\Big].
    \end{aligned}
  \end{gather}
We observe that
\[
p^m\dfrac{\d\r}{\d p}\subset \r
\]
and therefore
\[
 0=\oint_\gamma p^m\dfrac{\d}{\d p}\Big[\delta\Exp{\ad
 H}(\partial_{\mu n})\Big] p=\sum_{\mu=0}^M\oint_{\gamma_\mu} p^m\dfrac{\d}{\d p}\Big[\delta\Exp{\ad
 H}(\partial_{\mu n})\Big].
\]
From \eqref{t000n001} we derive
\begin{multline}\label{uf}
0=\oint_{\gamma_0}p^m\dfrac{\d}{\d p}\Big[\delta\Exp{\ad
H_{0>}}(\partial_{\mu n})\Big]\d p+\oint_{\Gamma_0}
     p(z_0)^m(n\delta_{\mu 0 }z_0^{n-1}-(1-\delta_{\mu 0})\delta_{n0}z_0^{-1})\d
     z_0\\+\sum_{i=1}^M\Big(
\oint_{\gamma_i}p^m\dfrac{\d}{\d p}\Big[\delta\Exp{\ad
H_{i}^-}(\partial_{\mu n})\Big]\d p+\delta_{\mu
i}\oint_{\Gamma_i}p(z_i)^m(n z_i^{n-1}+ \delta_{n0}z_i^{-1})\d z_i
     \Big).
\end{multline}
Therefore, recalling \eqref{phi} and
\[
p^m\dfrac{\d\g_i^-}{\d p}\subset \g_i^-
\]
we may write \eqref{uf} as follows
\[
m\Phi_{\mu n,m}=\frac{1}{2\pi\I}\oint_{\Gamma_\mu} p(z_\mu)^mn
z_\mu^{n-1}\d z_\mu
     + \delta_{n0}\frac{1}{2\pi\I}\Big(\oint_{\Gamma_\mu}\frac{p(z_\mu)^m}{z_\mu}\d
z_\mu-\oint_{\Gamma_0}\frac{p(z_0)^m}{z_0}\d z_0\Big).
\]
and \eqref{inversion}  implies
\begin{gather}\label{muchas}
m\Phi_{\mu n,m}=\frac{1}{2\pi\I}\oint_{\Gamma_\mu} p(z_\mu)^mn
z_\mu^{n-1}\d z_\mu
     + (1-\delta_{\mu 0})\delta_{n0}\Big(q_\mu^m-\frac{1}{2\pi\I}\oint_{\Gamma_0}\frac{p(z_0)^m}{z_0}\d
z_0\Big).
\end{gather}
where it must be understood that when $\mu=0$ the second term of the
r.h.s. vanishes. Hence, as
\[
\log\Big(1-\frac{q}{p}\Big)=-\sum_{m=1}^\infty\frac{1}{m}\frac{q^m}{p^m},\quad
\Big|\frac{q}{p}\Big|>1,
\]
we immediately derive \eqref{H0>}.
%
\end{proof}

As a byproduct of the above proof we get

\begin{cor}\label{potentialh01}
   The following relation
  \begin{gather}\label{h01z}
\sigma_{\mu n}=-\frac{1}{n+(1-\delta_{\mu 0})
\delta_{n0}}\pde{(\log\tau)_x}{t_{\mu n}},\quad
\sigma_{01}=-(\log\tau)_{xx}
  \end{gather}
  holds.
\end{cor}
\begin{proof}
  We prove the Theorem in the following steps:
  \begin{enumerate}
    \item
If we put $m=1$ in \eqref{muchas} we get
%
\begin{gather}\label{hhh}
\pde{h_{01}}{t_{\mu n}}=\frac{1}{2\pi\I}\oint_{\Gamma_\mu}
(nz_\mu^{n-1}+(1-\delta_{\mu 0}) \delta_{n0}z_\mu^{-1})p(z_\mu)\d
z_\mu,
\end{gather}
where we have 
taken
into account that
\[
\oint_{\Gamma_0}p(z_0)z_0^{-1}\d z_0=0.
\]
 \item  We use the inversion formula \eqref{inversion} in \eqref{hhh}
 we get
\begin{gather}\label{hhhh}
\pde{h_{01}}{t_{\mu
n}}=\sum_{l=-1,0,1,\dots}\frac{1}{2\pi\I}\oint_{\Gamma_\mu}
(nz_\mu^{n-1}+(1-\delta_{\mu 0}) \delta_{n0}z_\mu^{-1})\sigma_{\mu
l}\d z_\mu,
\end{gather}
and the desired result follows at once.
\item From the identity
\[
\delta\Exp{\ad H_0^-}\big(\pde{}{x}\big)=-\Exp{\ad
H_0^-}(p)-p=-z_0-p
\]
we get
\[
\pde{h_{01}}{x}=\frac{1}{2\pi\I}\oint_{\gamma_0}p\frac{\d z_0}{p}\d
p=\frac{1}{2\pi\I}\oint_{\Gamma_0}p(z_0)\d z_0=\sigma_{01}
\]
 \end{enumerate}
\end{proof}

Observe that all the coefficients $\sigma_{\mu n}$ are determined in
terms of $h_{01}$ and its time derivatives. Moreover, as all the
coefficients $d_{\mu n}$ are rational functions of the $\sigma_{\mu
m}$, for example:
\begin{align*}
d_{01}&=-\sigma_{01},&d_{02}&=-\sigma_{02},&d_{03}&=-\sigma_{03}+\sigma_{01}^2,\\
d_{i\,-1}&=\sigma_{i1},& d_{i0}&=\frac{\sigma_{i2}}{\sigma_{i1}},&
d_{i1}&=\frac{\sigma_{i3}\sigma_{i1}-\sigma_{i2}^2}{\sigma_{i1}^3},
\end{align*}
all the Lax functions may be written in terms of $h_{01}$ and its
$t$-derivatives.

Finally, we may write the contents of Theorem \ref{cauchy-pro} as
follows
\begin{cor}
The following identity holds
\begin{multline*}
\Big[\delta\Exp{\ad H_{0>}}(\partial_{\mu
n})\Big](p)=-\frac{1}{2\pi\I}\oint_{\gamma_\mu}
\log\Big(1-\frac{q}{p}\Big)\frac{\d z_\mu^n}{\d q}\d
q\\+(1-\delta_{\mu 0})
\delta_{n0}\Big(\log\Big(1-\frac{q_\mu}{p}\Big)-\frac{1}{2\pi\I}\oint_{\gamma_0}\log\Big(1-\frac{q}{p}\Big)\frac{\d\log(z_0(q))}{\d
q}\d q\Big),\quad p\in \bar\C\backslash D_0.
\end{multline*}
\end{cor}

For example, if we exclude the times $t_{i0}$ from the discussion we
get the suggesting formula
\[
 \Big[\delta\Exp{\ad H_{0>}}(\partial_{\mu
n})\Big](p)=-\frac{1}{2\pi\I}\oint_{\gamma_\mu}
\log\Big(1-\frac{q}{p}\Big)\frac{\d z_\mu^n}{\d q}\d q
\]

\subsection{Undressing solutions of the Whitham hierarchy}

In \S \ref{dressing} we have proved that the differential version of
the factorization problem \eqref{factorization2} may be described in
terms of the Whitham hierarchy. Here we show the equivalence between
both descriptions by  proving that any solution of the Whitham
hierarchy may be formally undressed; i.e., it comes from a
convenient factorization problem.

\begin{teh}\label{undressing}
Any set of Lax functions $z_\mu$ and zero-curvature functions
$\Omega_{\mu m}$ as in  \eqref{lax_functions}-\eqref{def-omega}
satisfying the Whitham hierarchy \eqref{lax-whitham},  may be
obtained by a dressing procedure based in the factorization
 problem \eqref{factorization2} as described in Theorem \ref{Theorem-fac-whitham}.
 \end{teh}
\begin{proof}
If we take as given the complex numbers $q^{(0)}_i$ and the
functions $q_i,d_{\mu n}$ from Proposition \ref{fac-coefficients}
we may determine the coefficients $X_i$ and $h_{\mu n}$ up to
$x$-independent terms. This last fact is clear from the
construction of $z_\mu$ as a dressing of $p_\mu$. Indeed, we have
that $\Exp{\ad H^-_\mu}p_\mu:=\Exp{\ad \tilde H^-_\mu}\Exp{\ad
f_\mu(p)}p_\mu=\Exp{\ad \tilde H^-_\mu}p_\mu$, where
$f_\mu\in\mathfrak c_\mu^-$.

We now undress, using the canonical transformation $\Exp{-\ad \tilde
H^-_\mu}$, the Lax functions and zero-curvature forms: $z_\mu\to
p_\mu$ and $\Omega_{\mu n}\to\Omega^0_{\mu n}$ with
\begin{gather}\label{Omega0}
\Omega_{\mu n}^{0}=\delta\Exp{-\ad \tilde H_\mu^-}(\partial_{\mu
n})+\Exp{-\ad \tilde H_\mu^-}\Omega_{\mu n}.
\end{gather}
Then,
\begin{gather}\label{undressz}
 0=\partial_{\mu n}p_\nu=\{\Omega_{\mu n}^{0},p_\nu\}
\end{gather}
and
\begin{gather}\label{undressOmega}
 \pde{\Omega_{\mu n}^{0}}{t_{\nu l}}-\pde{\Omega_{\nu
l}^{0}}{t_{\mu n}}+\{\Omega_{\mu n}^{0},\Omega_{\nu l}^{0}\}=0.
\end{gather}
From \eqref{undressz} we deduce that
\begin{gather}
  \label{Omega0x}
\Omega_{\mu n,x}^{0}=0\end{gather}
 so that \eqref{undressOmega} implies
\begin{gather}
  \label{Omega0t}\pde{\Omega_{\mu n}^{0}}{t_{\nu l}}=\pde{\Omega_{\nu
l}^{0}}{t_{\mu n}}
\end{gather}
Moreover, for $n>0$ we have
\[
\Omega_{\mu n}-z^n_\mu\in\g_\mu^-.
\]
Thus, $\Exp{-\ad H_\mu^-}\Omega_{\mu n}-p_\mu^n\in\g_\mu^-$ and
\eqref{Omega0} and \eqref{Omega0x} allow us to  deduce
\begin{gather*}
 \Omega^0_{\mu n}-p_\mu^n\in\mathfrak c_\mu^-\subset\g_\mu^-,\\
 \Omega^0_{i 0}+\log(p-q^{(0)}_i)\in\mathfrak c_i^-\subset\g_i^-.
\end{gather*}
Hence, recalling \eqref{Omega0t} we get
\[
\Omega_{\mu n}^0=\pde{(T_\mu+f_\mu)}{t_{\mu n}},\quad \text{ for
some }f_\mu\in\mathfrak c_\mu^-,
\]
and we can write
\[
\delta \Exp{-\ad f_\mu}(\partial_{\mu m})+\Exp{-\ad
f_\mu}\Omega_{\mu m}^0=\delta \Exp{\ad T_\mu}(\partial_{\mu m}).
\]
Therefore, if
\[
H_\mu^-=C(\tilde H_\mu^-,f_\mu)\in\g_\mu^-\text{; i.e., $\Exp{ \ad
H_\mu^-}=\Exp{\ad \tilde H_\mu^-}\Exp{\ad f_\mu}$},
\]
where $C$ was introduced in \eqref{campbell}, we have
\begin{gather}\label{Omegaundress}
\Omega_{\mu n}=\delta\Exp{\ad  H_\mu^-}(\partial_{\mu n})+\Exp{\ad
 H_\mu^-}\delta \Exp{\ad T_\mu}(\partial_{\mu
n})=\delta(\Exp{\ad  H_\mu^-}\Exp{\ad T_\mu})(\partial_{\mu n}),
\quad z_\mu=\Exp{\ad  H_\mu^-}\Exp{\ad T_\mu}p_\mu.
\end{gather}

Finally, from definition the zero-curvature connection $\Omega_{\mu
n}\in\r$ and there locally exists $H\in\r$ such that
\begin{gather}\label{jj}
  \Omega_{\mu n}=\delta\Exp{\ad  H}(\partial_{\mu n}),
\end{gather}
so that \eqref{Omegaundress} and \eqref{jj} leads us to the
factorization \eqref{factorization2} for some $G_\mu$.
\end{proof}

\subsection{Dressing transformations for the Whitham hierarchy}

In this section we show how to dress any solution of the Whitham
hierarchy by using the factorization problem technique. Let
$z^{(1)}$ be  Lax functions as described in \eqref{lax_functions},
with coefficients denoted by $q^{(1)}_i$ and $d^{(1)}_{\mu m}$, and
$\Omega^{(1)}_{\mu m}$, as defined in \eqref{def-omega}, so that the
Whitham hierarchy \eqref{lax-whitham} is satisfied:
\[
\pde{z^{(1)}_\nu}{t_{\mu n}}=\{\Omega^{(1)}_{\mu n},z^{(1)}_\nu\}.
\]
Let us assume that $q^{(1)}_i\in D_i$ so that there exists a
Hamiltonian $H^{(1)}\in\r$ with
\[
\Omega_{\mu n}^{(1)}=\delta\Exp{\ad H^{(1)}}(\partial_{\mu n}).
\]

Given  new \emph{initial conditions} $G_\mu$, $\mu=0,1,\dots,M$, the
factorization problem
\begin{gather}\label{factorization4}
\Exp{\ad H^{(1)}}\Exp{\ad G_\mu}=\Exp{-\ad H_\mu^-} \Exp{\ad
H^{(2)}},\quad H_\mu^-\in\g_\mu^-,\;H^{(2)}\in\r,
\end{gather}
will lead to a dressing procedure of the solution $z^{(1)}_\mu$ of
the Whitham hierachy as described below
\begin{pro}
  The new Lax functions
  \[
z^{(2)}_\mu=\Exp{\ad H_\mu^-}z^{(1)}_\mu
  \]
  are of the form \eqref{lax_functions} with new coefficients
  $q^{(2)}_i$ and $d_{\mu l}^{(2)}$ determined by $H_\mu^-$. The
  functions
  \[
  \Omega_{\mu m}^{(2)}=\begin{cases}
    \big(\big(z_\mu^{(2)}\big)^n\big)_{(\mu,+)},& n>\delta_{\mu 0},\\[8pt]
    -\log(p-q^{(2)}_i),& n=0,\mu=i=1,\dots,M,
  \end{cases}
  \]
  (in this case $(\cdot)_{(i,+)}$ projects in the span $\{\log(p-
q_i^{(2)}),(p- q_i^{(2)})^{-n}\}_{n=1}^\infty$ and $(\cdot)_{(0,+)}$
onto the span of $\{p^m\}_{m=0}^\infty$) have zero-curvature.
Moreover, the Whitham hierarchy
\[
\pde{z_\nu^{(2)}}{t_{\mu m}}=\{\Omega_{\mu m}^{(2)},z_\nu^{(2)}\}
\]
is satisfied.
\end{pro}
\begin{proof}
We take right logarithmic derivative of \eqref{factorization4} to
obtain
\begin{gather}\label{delta}
 \delta\Exp{\ad H_\mu^-}(\partial_{\nu n})+\Exp{\ad
H_\mu^-}(\Omega_{\nu n}^{(1)})= \delta\Exp{\ad
H^{(2)}}(\partial_{\nu n})=:\Omega_{\nu n}^{(2)}.
\end{gather}
 As $\Omega_{\nu n}^{(1)}$ is
holomorphic in $D_\mu$, for all $\mu\neq \nu$, we deduce that
$\Omega^{(2)}_{\nu n}$ is also holomorphic in $D_\mu$, $\forall
\mu\neq\nu$. When $\mu=\nu$ we have a singular behavior at
$p=q_\nu^{(1)}$ and we obtain $\Omega^{(2)}_{\nu n}$ with the same
structure as in \eqref{omega1}. If we write the factor $\Exp{\ad
H_\mu^-}$ as in Proposition \ref{fac-coefficients} and $X_i$ is
defined by
\[
\int_x^{X_i}\frac{\d x}{h_{i1}(x)}=1,
\]
we get, for example, the following coefficients of $z^{(2)}_\mu$:
\begin{align*}
 q_i^{(2)}&=X_{i,x} q_i^{(1)}\big|_{x=X_i}+h_{i0},\\
 d_{i\,-1}^{(2)}&= d_{i\,-1}^{(1)}|_{x=X_i}X_{i,x},\\
d_{i0}^{(2)}&=(d_{i0}^{(1)}+h_{i2,x}d^{(1)}_{i\,-1}+2h_{i2}d^{(1)}_{i\,-1,x})|_{x=X_i},\\
d_{01}^{(2)}&=d_{01}^{(1)}-h_{01,x}.
\end{align*}
 Moreover, the analysis of \eqref{delta} leads to the proof of
all the other properties. For example, from $\Omega_{\mu
n}^{(2)}=\delta\Exp{\ad H^{(2)}}(\partial_{\mu n})$ we deduce the
zero curvature condition for the $\{\Omega_{\mu n}^{(2)}\}$.
\end{proof}
Now,  we introduce
 \[
  H^{(0)}:=T(p)\in\mathcal \r,\quad T:=\sum_{\mu=0}^M T_\mu(p)
  \]
  for which
\begin{gather}\label{omega1}
  \Omega_{\mu n}^{(0)}:=\delta\Exp{\ad H^{(0)}}(\partial_{\mu
m})=\begin{cases}
    p^n,& \mu=0,\\[5pt]
    -\log(p- q_i^{(0)}),&\mu=i,\quad n=0,\\[5pt]
    \dfrac{1}{(p-q_i^{(0)})^m},&\mu=i,\quad n\geq 1,
  \end{cases}
\end{gather}
for this reason we say that $\Exp{\ad H^{(0)}}$ is a vacuum solution
of the Whitham  hierarchy. Indeed, its dressing
\begin{gather*}\label{fac0}
\Exp{\ad H^{(0)}}\Exp{\ad G_\mu}=\Exp{-\ad\tilde H_\mu^-}\Exp{\ad
H^{(1)}},\quad \tilde H_\mu^-\in\g_\mu^-,\;H^{(1)}\in\r
\end{gather*}
---giving $H^{(1)}$ and a new solution $\{\Omega_{\mu n}^{(1)}\}$ of
the Whitham hierarchy--- is just the factorization problem
\eqref{factorization2} when we replace
\[
\Exp{\ad \tilde H_\mu^-}\Exp{\ad(\sum_{\nu\neq
\mu}t_{\nu})}=\Exp{\ad H_\mu^-},\quad H_\mu^-\in\g_\mu^-.
\]

\section{String equations in the Whitham hierarchy}

In this section we study the formulation of the Whitham hierarchy in
terms of twistor or string equations and the relation of this
formulation with the dressing method described above. We first
introduce the Orlov--Schulman operators for the Whitham hierarchy in
terms of the factorization problem and then obtain the string
equation formulation as a consequence of the factorization problem.
In the forthcoming paper \cite{futuro} we will show that, in fact,
the string equations give all solutions of the Whitham hierarchy.
Then, string equations and factorization problem are equivalent
tools to formulate the Whitham hierarchy. Finally, we introduce a
very special class of string equation whose construction is based on
centerless Virasoro algebra within the Hamiltonian functions, and
therefore we refer to this  as the Virasoro class of string
equations.

\subsection{Lax and Orlov--Schulman functions of the Whitham hierarchy}

The Lax functions  \eqref{laxf} may be written as
\begin{align*}
  z_\mu&=\Exp{\ad H_\mu^-}\Exp{\ad T_\mu}p_\mu,\quad
  \mu=0,1,\dots,M.
 \end{align*}

Observe that  if we define $(p_\mu(p),x_\mu(x,p))$  by
\begin{gather}\label{pxm} (p_\mu,x_\mu):=\begin{cases} (p,x),
&\mu=0,\\[5pt]\Big((p-q_i^{(0)})^{-1},-x(p-q_i^{(0)})^2\Big), &
\mu=i\in\s.
\end{cases}
\end{gather}
we have
\[
\{p_\mu,x_\mu\}=1.
\]
In terms of $x_\mu$ the Orlov--Schulman function $m_\mu$ is
defined as follows
\begin{gather}\label{orlov}
m_\mu:=\Exp{\ad H_\mu^-}\Exp{\ad T_\mu}x_\mu,
\end{gather}
so that is canonically conjugated to $z_\mu$; i.e,
\[
\{z_\mu,m_\mu\}=1.
\]
Notice that the quasi-classical Lax equations also hold for the
Orlov--Schulman functions:
\begin{align}\label{orlov-whitham}
\pde{m_\nu}{ t_{\mu n}}&=\{\Omega_{\mu n},m_\nu\}.
\end{align}

We  now give a closer look to these functions
\begin{pro}
The Orlov--Schulman functions have the form
\begin{gather}\label{orlov-schulman} m_\mu=\sum_{n=1}^{\infty}nt_{\mu
n}z_\mu^{n-1}+\frac{t_{\mu0}}{z_\mu}+ \sum_{n\geq 2}v_{\mu
n}z_\mu^{-n}, \text{ where $t_{01}:=x$}
\end{gather}
and
\[
v_{\mu\; n+1}=\begin{cases}   -X_i, & \mu=i=1,\dots,M,\; n=0,\\
-(nh_{in}+ g_{in}(h_{i\,n-1},\dots,h_{i2}))\big|_{x=X_i},&\mu=i=1,\dots,M,\; n>0,\\
-(nh_{0n}+ g_{0n}(h_{0\, n-1},\dots,h_{01})), & \mu =0,\; n\geq 0,
\end{cases}
\]
being $g_{\mu n}$  differential polynomials.
\end{pro}
\begin{proof}
From \eqref{orlov} we deduce that
\begin{align*}
  m_\mu=\Exp{\ad H_\mu^-}\Big(x_\mu+\frac{\partial T_\mu}{\partial p_\mu}\Big),
\end{align*}
so that
\begin{align*}
  m_\mu&=\Exp{\ad H_\mu^-}x_\mu+(1-\delta_{\mu 0})t_{\mu
0}z_\mu^{-1}+\sum_{n=1+\delta_{\mu 0}}^\infty n t_{\mu n}z_\mu^{n-1}
\end{align*}
Now, we evaluate
\begin{gather}\label{adpsix}
\begin{aligned}
\Exp{\ad H_i^-}x_i&=-\big(X_i+\sum_{n=2}^\infty(nh_{in}+\tilde
g_{in}(h_{i\,n-1},\dots,h_{i2}))\big|_{x=X_i}
\Big(\frac{p-q_i}{X_{i,x}}\Big)^{n-1}\big)z_i^{-2},\\
 \Exp{\ad H_0^-}x&=x+t_{00}p^{-1}-\sum_{n=1}^\infty(nh_{0n}+ \tilde g_{0n}(h_{0\,
 n-1},\dots,h_{01}))p^{-n-1}
\end{aligned}
\end{gather}
where $\tilde g_{\mu n}$ are differential polynomials, but as
\begin{align*}
  \frac{p-q_i}{X_{i,x}}&=z_i^{-1}+h_{i1}|_{x=X_i}z_i^{-2}+O(z_i^{-3}),\\
  p^{-1}&=z_0^{-1}+h_{01}'z_0^{-3}+O(z_0^{-4})
\end{align*}
we get \eqref{orlov-schulman}.
\end{proof}
Observe that the first coefficients of $m_\mu$ are
\begin{align*}
  v_{i2}&=-X_i,\\
  v_{i3}&=-2h_{i,2}|_{x=X_i},\\
  v_{02}&=-h_{01}\\
  v_{03}&=-2h_{02}.
\end{align*}


\subsection{The factorization problem and strings equations}
Let us  define  new canonical pairs $(\hat z_\mu,\hat m_\mu)$ and
$(\hat P_\mu,\hat Q_\mu)$ given by
\begin{gather}\label{hateando}
\begin{aligned}
  \hat z_\mu&:=\Exp{\ad H_\mu^-}\Exp{\ad T_\mu}p,\\
  \hat m_\mu&:=\Exp{\ad H_\mu^-}\Exp{\ad T_\mu}x,
\end{aligned}\quad
\begin{aligned}
  \hat P_\mu&:=\Exp{\ad G_\mu}p,\\
  \hat Q_\mu&:=\Exp{\ad G_\mu}x.
\end{aligned}
\end{gather}
Observe that
\begin{align*}
z_\mu&=p_\mu(\hat z_\mu),\\
m_\mu&=x_\mu(\hat m_\mu,\hat z_\mu),
\end{align*}
where the functions are defined in \eqref{pxm}.

Now, we are ready to give a first version of the string or twistor
equations for the Whitham hierarchy:
\begin{pro}
For any given  solution of the factorization problem
\eqref{factorization2} with associated canonical pairs $(\hat
z_\mu,\hat m_\mu)$, $(\hat P_\mu,\hat Q_\mu)$, as defined in
\eqref{hateando},  the following string equations hold
\begin{gather}\label{hattwistor}
\begin{aligned}
\hat P_{\nu}(\hat z_{\nu}, \hat m_\nu)&=\hat P_\mu( \hat z_\mu, \hat m_\mu)\in \r,\\
 \hat Q_\nu(\hat z_\nu,\hat m_\nu)&=Q_\mu(\hat z_\mu,\hat m_\mu)\in
 \r.
\end{aligned}
\end{gather}
\end{pro}
\begin{proof}
The factorization \eqref{factorization2} implies
\begin{gather}\label{intwistor}
\begin{aligned}
  \hat P_\mu(\hat z_\mu,\hat m_\mu)&=\Exp{\ad H_\mu^-}\Exp{\ad T_\mu}\Exp{\ad G_\mu }p=\Exp{\ad H}p=\Pi,\\
 \hat Q_\mu(\hat z_\mu,\hat m_\mu)&=\Exp{\ad H_\mu^-}\Exp{\ad T_\mu}\Exp{\ad G_\mu }x=\Exp{\ad
 H}x=\Theta.
\end{aligned}
\end{gather}
Notice that
\begin{gather}\label{hatphi}
\hat \phi_\mu(p,x):=(\hat P_\mu(p,x),\hat Q_\mu(p,x))
\end{gather}
is a canonical transformation; i.e.,
\[
\{\hat P_\mu,\hat Q_\mu\}=1,
\]
 that together with
\eqref{intwistor} ensures that
\begin{gather}\label{hatchart}
\hat\phi_\mu(\hat z_\mu,\hat m_\mu)=\hat\phi_\nu(\hat z_\nu,\hat
m_\nu)=(\Pi,\Theta),
\end{gather}
and \eqref{hattwistor} follows.
\end{proof}

The string equations \eqref{hatchart} have an interesting
interpretation in terms of transition functions between different
canonical pairs
\begin{gather}\label{chart}
(\hat z_\mu,\hat m_\mu)=\hat\phi_{\mu\nu}(\hat z_\nu,\hat
m_\nu),\quad \hat\phi_{\mu\nu}:=\hat\phi_\mu^{-1}\circ\hat\phi_\nu.
\end{gather}
Now, we define the canonical transformation
\[
\psi_\mu(p,x):=(p_\mu(p),x_\mu(p,x))
\]
in terms of which the associated solutions of the Whitham hierarchy
are
\[
(z_\mu,m_\mu)=\psi_\mu(\hat z_\mu,\hat m_\mu).
\]
We also introduce
\[
\phi_\mu=(P_\mu,Q_\mu):=\hat{\phi}_\mu\circ\psi_\mu^{-1},\quad
\psi_\mu^{-1}=(\pi_\mu,\theta_\mu)=\begin{cases}
(p^{-1}+q_i^{(0)},-p^2x),&\mu=i\in\s,
\\(p,x),&\mu=0,
\end{cases}
\]
so that
\begin{gather}\label{PQ}
P_\mu:=\hat P_\mu(\pi_\mu(p,x),\theta_\mu(p,x))
, \quad Q_\mu:=\hat Q_\mu(\pi_\mu(p,x),\theta_\mu(p,x))
\end{gather}
and
\[ \{P_\mu,Q_\mu\}=1.
\]
Observe that this definition is equivalent to
\[
P_\mu(p_\mu,x_\mu)=\hat P_\mu(p,x),\quad Q_\mu(p_\mu,x_\mu)=\hat
Q_\mu(p,x).
\]

 Then, the connection among
the different Lax and Orlov--Schulman functions are given by
\[
(z_\mu,m_\mu)=\phi_{\mu\nu}(z_\nu,m_\nu), \quad
\phi_{\mu\nu}:=\psi_\mu\circ\hat\phi_{\mu\nu}\circ\psi_\nu^{-1}=\phi_\mu^{-1}\circ\phi_\nu.
\]
Therefore,
\[
\phi_\mu(z_\mu,m_\mu)=\phi_\nu(z_\nu,m_\nu)=(\Pi,\Theta)
\]
and
\begin{pro}
Given a solution of \eqref{factorization2} and functions
$(P_\mu,Q_\mu)$ as defined in \eqref{PQ}  the string equations
\begin{gather}\label{twistor}
\begin{aligned}
P_{\nu}(z_{\nu}, m_\nu)&=P_\mu( z_\mu, m_\mu)\in \r,\\
 Q_\nu(z_\nu,m_\nu)&=Q_\mu(z_\mu,m_\mu)\in \r,
\end{aligned}
\end{gather}
hold  $\forall \mu,\nu=0,1,\dots,M$.
\end{pro}

Notice that new initial conditions $\tilde G_\mu$ of the form
\[
\Exp{\ad \tilde G_\mu}=\Exp{\ad G}\Exp{\ad G_\mu}\text{ or $\tilde
G_\mu=C(G,G_\mu)$,}
\]
lead to
\[
\tilde P_\mu=P(P_\mu,Q_\mu),\quad  \tilde Q_\mu=Q(P_\mu,Q_\mu).
\]
Thus, the corresponding string equations are constructed in terms of
the initial non-tilded ones.

\subsection{A special class of string equations related to a centreless  Virasoro algebra}\label{string}
Consider the Hamiltonian
\begin{gather}\label{gvir}
G^{(0)}_\mu=\frac{x}{\hat \xi'_\mu(p)},
\end{gather}
which generate the canonical transformation
\[
(p,x)\to(\hat f_\mu(p),x/\hat f'_\mu(p)), \quad \hat
f_\mu:=\hat\xi^{-1}_\mu(1+\hat \xi_\mu(p)).
\]
Observe that these Hamiltonians close a Lie subalgebra
$\mathfrak{vir} :=\{xf(p), f:\C\to\C\}$ as
$\{xf(p),xg(p)\}=x(f'(p)g(p)-f(p)g'(p))\subset\mathfrak{vir}$. In
fact, $\mathfrak{vir}$ is a centerless Virasoro algebra with
generators
\begin{gather}\label{vir-gen}
l_n:=xp^{n-1}
\end{gather}
satisfying
\[
\{l_n,l_m\}=(n-m)l_{n+m}.
\]
The functions $\hat\xi_\mu,\hat f_\mu$ corresponding to the Virasoro
generators \eqref{vir-gen} are
\[
\hat\xi_\mu=\begin{cases}  \dfrac{p^{2-n}}{2-n},& n\neq 2,\\ \log
p,& n=2,
\end{cases}\quad\hat
f_\mu =\begin{cases}
  (2-n)\Big(1+\dfrac{p^{2-n}}{2-n}\Big)^{\frac{1}{2-n}},& n\neq 2,\\
  \Exp{}p,& n=2.
\end{cases}
\]

We will also use the harmonic Hamiltonian
\[
  R:=\frac{1}{2}(p^2+x^2)
\]
 which generates the canonical transformation
\[
(p,x)\rightarrow (-x,p).
\]
Let us consider a splitting $\s=I\cup J$, $I\cap J=\emptyset$, and
define the initial conditions
\begin{gather}\label{initial}
\Exp{\ad G_\mu}:=\begin{cases}   \Exp{\ad G_i^{(0)}}\Exp{\ad  R},&
i\in I,\\
\Exp{\ad G_\mu^{(0)}},& \mu\in J\cup\{0\},
\end{cases}
\end{gather}
in terms $G^{(0)}_\mu$ as defined in  \eqref{gvir}. It is easy to
realize that
\begin{gather}\label{Eq}
\begin{aligned}
  (\hat P_0,\hat Q_0)&=\Big(\hat f_0(p),\frac{x}{\hat f'_0(p)}\Big),\\
  (\hat P_i,\hat Q_i)&=\begin{cases}
    \Big(-\dfrac{x}{\hat f'_i(p)},\hat f_i(p)\Big),& i\in I,\\
 \Big(\hat f_i(p),\dfrac{x}{\hat f'_i(p)}\Big),& i\in J,
  \end{cases}
\end{aligned}
\end{gather}
 and the corresponding string equations are
\begin{gather}\label{strings}
\begin{aligned}
\hat f_0(\hat z_0)&=-\frac{\hat m_i}{\hat f_i'(\hat z_i)}\in\r,&
\hat f_i(\hat
z_i)&=\frac{\hat m_0}{\hat f_0'(\hat z_0)}\in\r,& i\in I,\\
\hat f_0(\hat z_0)&=\hat f_i(\hat z_i)\in\r,& \frac{\hat m_i}{\hat
f_i'(\hat z_0)}&=\frac{\hat m_0}{\hat f_0'(\hat z_0)}\in\r,& i\in J.
\end{aligned}
\end{gather}

Taking into account the invariance described in \eqref{invariance}
we deduce that he string equations \eqref{strings} also appear for
the following set of initial conditions
\begin{gather}\label{initial2}
\Exp{\ad G_\mu}:=\begin{cases}   \Exp{\ad G_i^{(0)}},&
i\in I,\\
\Exp{\ad G_\mu^{(0)}}\Exp{-\ad  R},& \mu\in J\cup\{0\}.
\end{cases}
\end{gather}
where now
\begin{gather}\label{Eq2}
\begin{aligned}
  (\hat P_0,\hat Q_0)&=\Big(\frac{x}{\hat f'_0(p)},-\hat f_0(p)\Big),\\
  (\hat P_i,\hat Q_i)&=\begin{cases}
    \Big(\hat f_i(p),\dfrac{x}{\hat f'_i(p)}\Big),& i\in I,\\
 \Big(\dfrac{x}{\hat f'_i(p)},-\hat f_i(p)\Big),& i\in J.
  \end{cases}
\end{aligned}
\end{gather}

We introduce the functions $f_\mu$ subject to
\[
f_\mu(p_\mu)=\hat f_\mu (p) \Rightarrow \begin{cases}   f_0(p)=\hat
f_0(p),&\mu=0\\
f_i(p)=\hat f_i\Big(\dfrac{1}{p}+q_i^{(0)}\Big),&\mu=i\in\s
\end{cases}
\]
so that
\begin{align*}
  (P_0, Q_0)&=\Big(-\frac{x}{f'_0(p)},f_0(p)\Big),\\
  (P_i,Q_i)&=\begin{cases}
    \Big( f_i(p),-\dfrac{x}{f'_i(p)}\Big),& i\in I,\\
 \Big(-\dfrac{x}{ f'_i(p)},f_i(p)\Big),& i\in J.
  \end{cases}
\end{align*}


 Therefore,  we get the string
equations
\begin{gather}\label{strings2}
\begin{aligned}
 f_0( z_0)&=-\frac{m_i}{f_i'(z_i)}\in\r,&
f_i(z_i)&=\frac{m_0}{f_0'(z_0)}\in\r,& i\in I,\\
f_0( z_0)&= f_i( z_i)\in\r,& \frac{ m_i}{ f_i'( z_0)}&=\frac{m_0}{
f_0'(z_0)}\in\r,& i\in J.
\end{aligned}
\end{gather}

These string equations reduces to Krichever type of string equations
considered in \cite{krichever} for $J=\mathbb S$ and to the
Takasaki--Takebe type \cite{21} for  $J=\emptyset$.

\section{Additional symmetries for the Whitham hierarchy}

This section is devoted to the analysis of the additional or master
symmetries of the Whitham hierarchy. For that aim we we characterize
the additional symmetries in terms of deformations of the
factorization problem \eqref{factorization2}. We compute then some
explicit examples of additional symmetries leading to functional
symmetries of the generalized Benney  gas equations. Finally, we
study its action on Virasoro string equations.

\subsection{Deformation of the factorization problem  and additional symmetries}
\label{additional}

The treatment of functional symmetries of dispersionless hierarchies
as additional symmetries was first given in \cite{40} for the
dispersionless KP hierarchy. Then, its formulation as a deformation
of a factorization problem for the $r$-th dispersionless Toda
hierarchy was considered in \cite{manas2}.

In this section we allow each initial condition Hamiltonian $G_\mu$
to depend on a external parameter $s$
\[
G_\mu:=G_\mu(s).
\]
 Then, the factorization
problem \eqref{factorization2} also depends on $s$
\begin{gather}\label{factorization3}
\Exp{\ad T_\mu}\Exp{\ad G_\mu(s)}=\Exp{-\ad H_\mu^-(s)}\Exp{\ad
H(s)},\text{ with $H_\mu^-(s)\in\g_\mu^{-}$ and $H(s)\in\r$}.
\end{gather}

 Thus, we deduce that
\begin{teh}
Additional symmetries of the Whitham hierarchy are characterized by
functions $F_\mu(z_\mu,m_\mu)$ as follows
  \begin{align*} \pde{z_\nu}{s}&=-\pde{F_\nu}{m_\nu}+\sum_{\mu=0}^M\{
  (F_\mu(z_\mu,m_\mu))_{(\mu,+)},z_\nu\},\\
  \pde{m_\nu}{s}&=-\pde{F_\nu}{z_\nu}+\sum_{\mu=0}^M\{(F_\mu(z_\mu,m_\mu))_{(\mu,+)},m_\nu\}.
\end{align*}
\end{teh}
\begin{proof}
Taking the right logarithmic derivative of \eqref{factorization3}
with respect to $s$ we get
\begin{gather}\label{add}
\begin{aligned}
\delta\Exp{\ad H_\mu^-}\Big(\pde{}{s}\Big)+ F_\mu( z_\mu,
m_\mu)&=\delta\Exp{\ad H}\Big(\pde{}{s}\Big),
\end{aligned}
\end{gather}
where
\[
F_\mu(z_\mu,m_\mu)=\hat F_\mu(\hat z_\mu,\hat m_\mu),\quad \hat
F_\mu:=\delta\Exp{\ad G_\mu}\Big(\pde{}{s}\Big).
\]
Observe that from the splitting
\[
F_\mu=F_\mu^-+F,\quad F:=\sum_{\nu=0}^M F_{(\nu,+)}
\]
with
\[
F_\mu^-\in\g_\mu^-,\quad F\in\r,
\]
and from \eqref{add} we get  that
 \begin{gather}\label{addsymmgen}
\begin{aligned}
  \delta\Exp{\ad H_\mu^-}\Big(\pde{}{s}\Big)&=-F_\mu^-=F-F_\mu,\\
\delta\Exp{\ad H}\Big(\pde{}{s}\Big)&=F.
\end{aligned}
\end{gather}
Therefore, from
\[
\pde{z_\mu}{s}=\{
  \delta\Exp{\ad H_\mu^-}\Big(\pde{}{s}\Big),z_\mu\},\quad
\pde{m_\mu}{s}=\{
  \delta\Exp{\ad H_\mu^-}\Big(\pde{}{s}\Big),m_\mu\}
\]
we get the desired result.
\end{proof}

 An important reduction is given by $t_{\mu n}=0$ for $n>N_\mu$.
If we assume that
 \begin{equation}\label{F}
 F_\mu(z_\mu,m_\mu)=c_\mu\log
z_\mu+\sum_{i,j\in\mathbb{Z}} c_{\mu,ij}z_\mu^im_\mu^j,
\end{equation}
and
\begin{gather}\label{orlov-schulman-cortado}
m_\mu=\sum_{n=1}^{N_\mu}nt_{\mu
n}z_\mu^{n-1}+\frac{t_{\mu0}}{z_\mu}+ \sum_{n\geq 2}v_{\mu
n}z_\mu^{-n},
\end{gather}
 imposing  $F_\mu(z_\mu,m_\mu)$ to have no terms proportional to
$z^n_\mu$ for $n>N_\mu$, we ensure  that the constraints  are
preserve. We request this for each of the products $z_\mu^im_\mu^j$:
\begin{align*}
z_\mu^im_\mu^j&=z_\mu^i(N_\mu t_{\mu
N_\mu}z_\mu^{N_\mu-1}+\dots+t_{\mu 1}+t_{\mu 0}z_\mu^{-1}+v_{\mu
2}z_\mu^{-2}+v_{\mu 3}z_\mu^{-3}+\dots)^j\\
&=(N_\mu t_{\mu N_\mu})^jz_\mu^{i+(N_\mu-1)j}+\cdots\Rightarrow
c_{\mu,ij}=0 \text{ if } i+(N_\mu-1)j>N_\mu.
\end{align*}
Hence,
\begin{equation}\label{Fform}
F_\mu(z_\mu,m_\mu)=c_\mu\log z_\mu+\sum_{n=1}^{N_\mu}\alpha_{\mu
n}\Big(\frac{m_\mu}{N_\mu z_\mu^{N_\mu-1}}\Big)z_\mu^n
\end{equation}
with $\alpha_n$ being analytic functions.

Some times is convenient to consider that only one of the initial
conditions is deformed, say the $\alpha$-component:
\[
\pde{G_\mu}{s}=\delta_{\mu\alpha}\pde{G_\alpha}{s},\quad
\forall\mu=0,1,\dots,M.
\]
In this case we get the following symmetry equations
 \begin{gather}\label{addsymmred} \begin{aligned}
  \pde{z_\nu}{s}&=\delta_{\nu\alpha}\pde{F_\alpha}{m_\alpha}+\{(F_\alpha(z_\alpha,m_\alpha))_{(\mu,+)},z_\nu\},\\
  \pde{m_\nu}{s}&=-\delta_{\nu\alpha}\pde{F_\alpha}{z_\alpha}+\{(F_\alpha(z_\alpha,m_\alpha))_{(\mu,+)},m_\nu\}.
\end{aligned}
\end{gather}

\subsection{Action of additional symmetries on the potential function}

Observe that if we express $m_\mu=m_\mu(\bt,z_\mu)$ we get
$F_\mu(z_\mu,m_\mu(z_\mu))=:f_\mu(z_\mu)$. Then, inspired by Theorem
\ref{potentialh01}, we get

\begin{teh}
The following relation
  \begin{gather*}
\pde{(\log\tau)_x}{s}=-\frac{1}{2\pi\I}\sum_{\mu=0}^M\oint_{\Gamma_\mu}p(z_\mu)\frac{\d
f_\mu}{\d z_\mu}\d z_\mu,
  \end{gather*}
  holds.
\end{teh}
\begin{proof}
From \eqref{add} and
  \[
0=\oint_\gamma p\frac{\d}{\d p}\Big[\delta\Exp{\ad
H}\Big(\pde{}{s}\Big)\Big]\d p= \sum_{\nu=0}^M\oint_{\gamma_\nu}
p\frac{\d}{\d p}\Big[\delta\Exp{\ad H}\Big(\pde{}{s}\Big)\Big]\d p
  \]
we conclude that
\[
0=\sum_{\nu=0}^M\Big[\oint_{\gamma_\nu}
  p\frac{\d}{\d p}\Big[\delta\Exp{\ad
H_\nu^-}\Big(\pde{}{s}\Big)\Big]\d p+\oint_{\gamma_\nu}p\frac{\d}{\d
p}[f_\nu(z_\nu)]\d p\Big].
\]
But
\[
p\frac{\d}{\d p}\Big[\delta\Exp{\ad H_i^-}
  \Big(\pde{}{s}\Big)\Big]\in\g_i^-
  \]
is holomorphic in $D_i$ and
\[
  p\frac{\d}{\d p}\Big[\delta\Exp{\ad H_0^-}
\Big(\pde{}{s}\Big)\Big]=-\pde{h_{01}}{s}p^{-1}+O(p^{-2})+\pde{t_{00}}{s}z_0^{-1}p\frac{\d
z_0}{\d p},
\]
and the stated result follows.
\end{proof}

Let us assume the expansion
\[
f_\mu=\sum_{n=-\infty}^{\infty}f_{\mu n}z_\mu^{n}
\]
and perform a change of variables $p\to z_\mu$ to get
\begin{align*}
\pde{h_{01}}{s}&=\frac{1}{2\pi
\I}\sum_{\mu=0}^M\oint_{\Gamma_\mu}\Big(\sum_{n=-\infty}^{\infty}f_{\mu
n}nz_\mu^{n-1}\Big)\Big(\sum_{l=-1,0,1,\dots}\sigma_{\mu
l}z_\mu^{-l}\Big)\d z_\mu\\&=
\sum_{\substack{\mu=0,\dots,M\\n=-1,0,1,\dots}}nf_{\mu n}\sigma_{\mu
n}=-f_{0\, -1}+\sum_{\substack{\mu=0,\dots,M\\n\geq 1}}nf_{\mu
n}\sigma_{\mu n}
\end{align*}
Therefore, \eqref{hhhh} gives
\begin{gather}\label{clara}
\pde{h_{01}}{s}=-f_{0\, -1}+f_{01}\pde{h_{01}}{x}
+\sum_{\substack{\mu=0,\dots,M\\n\geq 1+\delta_{\mu 0}}} f_{\mu n}
\pde{h_{01}}{t_{\mu n}}
\end{gather}
This will be a linear PDE for $h_{01}$ if we ensure that the
coefficients $f_{\mu n}$ that the dependence on the functions
$v_{\mu n}$ is restricted to $v_{02}=-h_{01}$, i.e. do not depend on
rational functions of $h_{01}$ and its derivatives. This is the case
always for the time reduction $t_{\mu n}=0$ for $n>N_\mu$,
$\forall\mu=0,1,\dots,M$. For example, we take
\[
F_\mu=\alpha_\mu\Big(\frac{m_\mu}{N_\mu
z_\mu^{N_\mu-1}}\Big)z^n_\mu,\quad n=1,\dots,N_\mu.
\]
Then,
\[
\frac{m_\mu}{N_\mu z_\mu^{N_\mu-1}}=t_{\mu\,
N_\mu}+\frac{N_\mu-1}{N_\mu}t_{\mu\,N_\mu-1}z_\mu^{-1}+\dots+\frac{1}{N_\mu}
t_{\mu 0}
z_\mu^{-N_\mu}+\frac{1}{N_\mu}v_{\mu2}z_\mu^{-N_\mu-1}+\cdots,
\]
and therefore
\[
\alpha_\mu\Big(\frac{m_\mu}{N_\mu z_\mu^{N_\mu-1}}\Big)=A_{\mu
0}+A_{\mu 1}z_\mu^{-1}+\cdots
\]
with
\begin{gather}\label{david}
\begin{aligned}
  A_{\mu0}&=\alpha_\mu(t_{\mu\,N_\mu}),\\
  A_{\mu1}&=\alpha'_\mu(t_{\mu\,N_\mu})\frac{N_\mu-1}{N_\mu}t_{\mu\,N_\mu-1},\\
  A_{\mu2}&=\alpha'_\mu(t_{\mu\,N_\mu})\frac{N_\mu-2}{N_\mu}t_{\mu\,N_\mu-2}+
  \frac{1}{2}\alpha''_\mu(t_{\mu\,N_\mu})\frac{(N_\mu-1)^2}{N_\mu^2}t_{\mu\,N_\mu-1}^2,\\
  &\vdots\\
  A_{\mu\,N_\mu}&=\alpha'_\mu(t_{\mu\,N_\mu})\frac{t_{\mu0}}{N_\mu}+
\alpha''_\mu(t_{\mu\,N_\mu}) \frac{\sum'_{r+s=N_\mu}t_{\mu r}t_{\mu
s}}{N_\mu}+\dots+\alpha_\mu^{(N_\mu)}(t_{\mu\,N_\mu})\frac{(N_\mu-1)^{N_\mu}}{N_\mu!N_\mu^{N_\mu}}t_{\mu\,N_\mu-1}^{N_\mu},\\
A_{\mu\,N_\mu+1}&=\alpha'_\mu(t_{\mu\,N_\mu})\frac{v_{\mu2}}{N_\mu}+
\alpha''_\mu(t_{\mu\,N_\mu}) \frac{\sum'_{r+s=N_\mu+1}t_{\mu
r}t_{\mu
s}}{N_\mu}+\dots+\alpha_\mu^{(N_\mu+1)}(t_{\mu\,N_\mu})\frac{(N_\mu-1)^{N_\mu+1}}{(N_\mu+1)!N_\mu^{N_\mu+1}}
t_{\mu\,N_\mu-1}^{N_\mu+1},\\
&\vdots
\end{aligned}
\end{gather}
Here $\sum'$ means that if $r=s$ then we multiply this contribution
by  $1/2$. We see that all the coefficients $A_{\mu
0},\dots,A_{\mu\, N_\mu}$ do not depend on the functions $v_{\mu
2}$, for all the others the coefficients $v_{\mu n}$ contribute. In
particular, in $A_{0\,N_0+1}$ depends on $v_{02}$.

We have the formula
\[
f_{\mu m}=A_{\mu\, n-m},
\]
and \eqref{clara} reads

\begin{gather}\label{montse}
\pde{h_{01}}{s}=-f_{0\, -1}+f_{01}\pde{h_{01}}{x}
+\sum_{\substack{\mu=0,\dots,M\\ 1+\delta_{\mu 0}\leq m\leq n}}
A_{\mu n-m} \pde{h_{01}}{t_{\mu n}}
\end{gather}

For $n=1,\dots, N_\mu-1$  the coefficients $A$'s that appear in the
above equations do not depend on any $v$'s,  $v_{02}=-h_{01}$ and
for $\mu=0,n=N_0$ the coefficient $A_{0\,N_0+1}$ do depend linearly
on $v_{02}=-h_{01}$.

Notice that \eqref{montse} and \eqref{david} allows us to describe
the motion of the potential $h_{01}$ of the Whitham hierarchy under
additional symmetries via a  linear PDEs.

\subsection{Functional symmetries of the Boyer--Finley--Benney system}
Let us take $N_0=2$ and $N_i=1$, so that the involved times are
 $\{t_{i0}=:x_i,t_{j1}=:y_j, t_{02}=:t\}_{i,j=1}^M$ and the PDE' system is the one presented in \S\ref{Sistema}.
Now, we have
\begin{align*}
  m_0&=2tz_0+x+t_{00}z_0^{-1}+v_{02}z_0^{-2}+\cdots,\quad t_{00}=x_1+\cdots+x_M\\
  m_i&=y_i+x_iz_i^{-1}+v_{i2}z_i^{-2}+\cdots,
\end{align*}
so that\begin{align*}
\alpha_0\Big(\frac{m_0}{2z_0}\Big)&=\alpha_0(t)+\alpha'_0(t)\frac{x}{2}z_0^{-1}+\Big(\alpha_0'(t)\frac{t_{00}}{2}+
\alpha_0''(t)\frac{x^2}{8}\Big)z_0^{-2}+\cdots,\\
\alpha_i(m_i)&=\alpha_i(y_i)+\alpha'_i(y_i)x_iz_i^{-1}+\Big(\alpha_i'(y_i)v_{i2}+
\alpha_i''(y_i)\frac{x_i^2}{2}\Big)z_i^{-2}+\cdots,
\end{align*}
 We put
$C_\mu=0$ as these symmetries corresponds to the first flows
$\pde{}{x_i}$. Then, in the context of \eqref{addsymmred} we have
three different type of generators:
\begin{align*}
  F_0^{(1)}&=\alpha_0\Big(\frac{m_0}{2z_0}\Big)z_0=\alpha_0(t)z_0+\alpha_0'(t)\frac{x}{2}+
  \Big(\alpha_0'(t)\frac{t_{00}}{2}+\alpha_0''(t)\frac{x^2}{8}\Big)z_0^{-1}+\cdots,\\
  F_0^{(2)}&=\alpha_0\Big(\frac{m_0}{2z_0}\Big)z_0^2=\alpha_0(t)z_0^2+\alpha_0'(t)\frac{x}{2}z_0+\Big(\alpha_0'(t)\frac{t_{0
0}}{2}+\alpha_0''(t)\frac{x^2}{8}\Big)+\cdots,\\
  F_i&=\alpha_i(m_i)z_i=\alpha_i(y_i)z_i+\alpha_i'(y_i)x_iz_i^{-1}+
  \Big(\alpha_i'(y_i)v_{i2}+\alpha_i''(y_i)\frac{x_i^2}{2}\Big)z_i^{-2}+\cdots.\\
 \end{align*}
 Therefore,
\begin{align*}
  \pde{F_0^{(1)}}{m_0}&=\frac{1}{2}\alpha_0'\Big(\frac{m_0}{2z_0}\Big),\\
  \pde{F_0^{(2)}}{m_0}&=\frac{1}{2}\alpha'_0\Big(\frac{m_0}{2z_0}\Big)z_0,\\
  \pde{F_i}{m_i}&=\alpha'_i(m_i)z_i,
\end{align*}
and
\begin{align*}
  (F_0^{(1)})_{(0,+)}&=\alpha_0(t)\Omega_{01}+\alpha_{0}'(t)\frac{x}{2},\\
  (F_0^{(2)})_{(0,+)}&=\alpha_0(t)\Omega_{02}+\alpha_{0}'(t)\frac{x}{2}\Omega_{01}+\alpha_{0}'(t)\frac{t_{0
0}}{2}+\alpha_{0}''(t)\frac{x^2}{8},\\
(F_i)_{(i,+)}&=\alpha_i(y_i)\Omega_{i1}.
\end{align*}

Hence, the evolution of the Lax functions under these three types
of symmetries is characterized by the following  PDE' system
\begin{align}
\label{s01}S_0^{(1)}&:\begin{cases}
  \pde{z_0}{s_0^{(1)}}=\dfrac{1}{2}\alpha_0'\Big(\dfrac{m_0}{2z_0}\Big)+\alpha_0(t)\pde{z_0}{x}-
  \dfrac{1}{2}\alpha_{0}'(t)\pde{z_0}{p},\\[8pt]
\pde{z_i}{s_0^{(1)}}=\alpha'_0(t)\pde{z_i}{x}-
  \dfrac{1}{2}\alpha_{0}'(t)\pde{z_i}{p},
\end{cases}\\
\label{s02}S_0^{(2)}&:\begin{cases}
  \pde{z_0}{s_0^{(2)}}=\dfrac{1}{2}\alpha_0'\Big(\dfrac{m_0}{2z_0}\Big)z_0+\alpha_0(t)\pde{z_0}{t}+
  \dfrac{1}{2}\alpha_{0}'(t)x\pde{z_0}{x}-
  \dfrac{1}{2}\alpha_{0}'(t)p\pde{z_0}{p}-\alpha_{0}''(t)\dfrac{x}{4}\pde{z_0}{p},\\[8pt]
\pde{z_i}{s_0^{(2)}}=\alpha_0(t)\pde{z_i}{t}+
  \dfrac{1}{2}\alpha_{0}'(t)x\pde{z_i}{x}-
  \dfrac{1}{2}\alpha_{0}'(t)p\pde{z_i}{p}-\alpha_{0}''(t)\dfrac{x}{4}\pde{z_i}{p},
\end{cases}
\\
\label{si}S_i&:\begin{cases}
  \pde{z_0}{s_i}=\alpha_i(y_i)\pde{z_0}{y_i},&\\[8pt]
  \pde{z_j}{s_i}=\alpha_i(y_i)\pde{z_j}{y_i},&j\neq i\\[8pt]
\pde{z_i}{s_i}=\alpha'_i(m_i)z_i+\alpha_i(y_i)\pde{z_i}{y_i},&
\end{cases}
\end{align}
We now analyze how the dependent variables $\{w,v_i,q_i\}_{i=1}^M$
evolve under these symmetries
\begin{itemize}
  \item   The  $S^{(1)}_0$ equations \eqref{s01} implies a
  transformation that only involves the independent variables
  $(x,t)$ characterized by the following PDEs
\begin{align*}
\pde{w}{s_0^{(1)}}-\alpha_0(t)\pde{w}{x}+\dfrac{x}{4}\alpha_0''(t)=0,\\
\pde{v_i}{s_0^{(1)}}-\alpha_0(t)\pde{v_i}{x}=0,\\
\pde{q_i}{s_0^{(1)}}-\alpha_0(t)\pde{q_i}{x}+\dfrac{1}{2}\alpha_0'(t)=0,
\end{align*}
and the symmetry transformation is
\begin{align*}
w(x,t)&\to w(x+f(t),t)-\frac{f''(t)}{4}\Big(x+ \frac{f(t)}{2}\Big),\\
v_i(x,t)&\to v_i(x+f(t),t),\\
q_i(x,t)&\to q_i(x+f(t),t)-\frac{f'(t)}{2}
\end{align*}
with $f:=s^{(1)}_0\alpha_0$. For the potential $W$ this symmetry
reads
\[
W(x,t)\to
W(x+f(t),t)-\frac{f''(t)}{8}x(x+f(t))-\frac{f'(t)}{2}\sum_{i=1}^Mx_i.
\]

 \item   In this case the  $S^{(2)}_0$ equations \eqref{s02} implies a
  transformation characterized by the following PDEs
\begin{align*}
\pde{w}{s_0^{(2)}}-\alpha_0(t)\pde{w}{t}-\frac{1}{2}\alpha'_0(t)x\pde{w}{x}-\alpha'_0(t)w+
\dfrac{t_{00}}{4}\alpha_0''(t)+\dfrac{x^2}{16}\alpha_0'''(t)=0,\\
\pde{v_i}{s_0^{(2)}}-\alpha_0(t)\pde{v_i}{t}-\dfrac{1}{2}\alpha'_0(t)x\pde{v_i}{x}-\dfrac{1}{2}\alpha'_0(t)v_i=0,\\
\pde{q_i}{s_0^{(2)}}-\alpha_0(t)\pde{q_i}{t}-\dfrac{1}{2}\alpha'_0(t)x\pde{q_i}{x}-\dfrac{1}{2}\alpha'_0(t)q_i-\dfrac{1}{4}\alpha_0''(t)x=0,,
\end{align*}
and the symmetry transformation is
\begin{align*}
w(x,t)&\to
T'(t)w(\sqrt{T'(t)}x,T(t))-\frac{t_{00}}{4}\frac{T''(t)}{T'(t)}+\frac{1}{16}
\{T,t\}_{\text{S}} x^2,\\
v_i(x,t)&\to \sqrt{T'(t)}
v_i(\sqrt{T'(t)}x,T(t)),\\
q_i(x,t)&\to\sqrt{T'(t)}
q_i(\sqrt{T'(t)}x,T(t))+\frac{1}{4}\frac{T''(t)}{T'(t)}x
\end{align*}
with $T:=T(s^{(2)}_0,t)$ characterized by the following relation
\[
\int_t^T\frac{\d t}{\alpha_0(t)}=s^{(2)}_0,
\]
and we have used the Schwarztian derivative
\[
\{T,t\}_{\text{S}}:=\Big(\frac{T''(t)}{T'(t)}\Big)'
-\frac{1}{2}\Big(\frac{T''(t)}{T'(t)}\Big)^2=\frac{T'''(t)}{T'(t)}
-\frac{3}{2}\Big(\frac{T''(t)}{T'(t)}\Big)^2,
\]
which must be not confused with the Poisson bracket.

 For the potential $W$ this symmetry reads
\[
W(x,t)\to
\sqrt{T'(t)}W(\sqrt{T'(t)}x,T(t))+\frac{1}{4}\frac{T''(t)}{T'(t)}x\sum_{i=1}^Mx_i+\frac{1}{48}
\{T,t\}_{\text{S}}x^3.
\]

\item    The  $S_i$-symmetry characterized by equations \eqref{si} implies a
  transformation that only involves the independent variables
  $(x,y_i)$ as follows\begin{align*}
\pde{w}{s_i}-\alpha_i(y_i)\pde{w}{y_i}&=0,&\\
\pde{v_j}{s_i}-\alpha_i(y_i)\pde{v_j}{y_i}&=0,&j\neq i,\\
\pde{v_i}{s_i}-\alpha_i(y_i)\pde{v_i}{y_i}-\alpha_i'(y_i)v_i&=0,&\\
\pde{q_j}{s_i}-\alpha_i(y_i)\pde{q_j}{y_i}&=0,&\\
\end{align*}
Thus, if $Y_i(s_i,y_i)$ is defined by
\[
\int_{y_i}^{Y_i}\frac{\d y_i}{\alpha_i(y_i)}=s_i,
\]
then, we have
\begin{align*}
w(y_i)&\to w(Y_i(y_i)),&\\
v_j(y_i)&\to v_j(Y_i(y_i)),& j\neq i\\
v_i(y_i)&\to Y_i'(y_i) v_i(Y_i(y_i)),&\\ q_i(y_i)&\to
q_i(Y_i(y_i)),&
\end{align*}
which in terms of the potential $W$ reads
\[
W(y_i)\to W(Y_i(y_i)).
\]
\end{itemize}

If we put $M_0=1$; i.e, we not consider the $t$-flow, the
transformation is
\begin{align*}
  v_i&\to X'(x)v_i(X(x)),\\
  q_i&\to X'(x)q_i(X(x))+t_{00}\frac{X''}{X'}
\end{align*}
where
\[
\int_x^X\frac{\d x}{\alpha_0(x)}=s_0.
\]
That in potential form is
\[
W\to X'(x)W(X(x))+\frac{t_{00}^2}{2}\frac{X''}{X'}.
\]

This symmetry together with the $S_i$ symmetries described above
constitute the well-known  conformal symmetries of the extended
Boyer--Finley system. When the $t$-flow is plugged in, and the
extended Benney system appears, then this $x$-conformal symmetry
disappears.

Nevertheless these additional symmetries, to the knowledge of the
authors are not known for the generalized Benney system
\cite{benney}
\begin{gather}\label{benney}
\begin{aligned}
\pde{q}{t}&=\pde{(q^2-2w)}{x},\\
\pde{v}{t}&=2\pde{(qv)}{x},\\
\pde{w}{y}&=\pde{v}{x}.
\end{aligned}
\end{gather}
In fact, we have proven
\begin{pro}
Given any three functions $Y(y),f(t),T(t)$ and a solution
$(w,q,v)$ of \eqref{benney}, then we have a new solution $(\tilde
w,\tilde q,\tilde v)$ given by
\begin{align*}
\tilde w&=
T'(t)w(\sqrt{T'(t)}(x+f(t)),Y(y),T(t))-\frac{t_{00}}{4}\frac{T''(t)}{T'(t)}+\frac{1}{16}
\{T,t\}_{\text{S}} (x+f(t))^2-\frac{f''(t)}{4}\Big(x+ \frac{f(t)}{2}\Big),\\
\tilde q&=\sqrt{T'(t)}
q(\sqrt{T'(t)}(x+f(t)),Y(y),T(t))+\frac{1}{4}\frac{T''(t)}{T'(t)}(x+f(t))-\frac{f'(t)}{2},\\
\tilde v&= \sqrt{T'(t)} Y'(y)v(\sqrt{T'(t)}(x+f(t)),Y(y),T(t)).
\end{align*}
\end{pro}

We must notice that the above functional symmetries do not respect
the shallow water reduction that appears in the limit $x=-y$.

\subsection{ Additional symmetries of Virasoro type and its action on string equations}

As we have seen  in \S\ref{additional} additional symmetries appears
when deformations of the initial conditions are considered. Here we
will consider initial conditions as in \eqref{initial} and
\eqref{gvir} with $G^{(0)}_\mu$ depending on a $s$ parameter as
follows:
\begin{gather}\label{curve-vira}
G^{(0)}_\mu=\frac{x}{\hat\xi'_\mu(p,s)},
\end{gather}
so that in the string equations \eqref{strings2}  we will have
functions
\[
f_\mu=f_\mu(s).
\]

Notice that \eqref{curve-vira} describes a curve in the Virasoro
algebra $\mathfrak{vir}$, and therefore describes the more general
motion for the set of initial conditions $G_\mu$.

The right logarithmic derivative of the initial conditions
\eqref{initial} with respect to the additional parameter $s$ is
\[
\delta\Exp{\ad
G_\mu}\Big(\pde{}{s}\Big)=\beta_\mu(p_\mu)x_\mu,,\quad
\beta_\mu:=-\frac{f_{s}(p_\mu,s)}{f_{p_\mu}(p_\mu,s)},
\]
and the corresponding additional symmetry generator is
\[
F_\mu=\beta_\mu( z_\mu) m_\mu
\]
so that
\[
\pde{z_\nu}{s}=\beta_\nu(z_\nu)+\sum_{\mu=0}^M\{
F_{(\mu,+)},z_\nu\}.
\]

Now, if we freeze times $t_{\mu n}=0$ for $n>N_\mu$ so that
\[
m_\mu=\sum_{n=1}^{N_\mu}n t_{\mu n}z_\mu^{n-1}+t_{\mu
0}z_\mu^{-1}+\sum_{n=2}^\infty v_{\mu n}z_\mu^{-n}
\]
and we require  the additional symmetry to leave those times
invariant, we must have
\[
\beta_\mu(z_\mu)=\sum_{l=- 1}^{\infty}b_{\mu l}z_\mu^{-l}.
\]

Let us take, for simplicity, Virasoro type generators
\[
\beta_\nu=c_\nu z_\nu^{1-n_\nu},\quad n_\nu=1,\dots,N_\nu,\quad
c_\nu\in\C
\]
 so that
\[
\pde{z_\nu}{s}=c_\nu z_\nu^{1-n_\mu}+
\sum_{\substack{\mu=0,\dots,M\\n=1,\dots,N_\mu-n_\mu}}
(n+n_\mu)t_{\mu \,n+n_\mu}c_\mu\pde{z_\nu}{t_{\mu n}}.
\]
whose integration,  leads to
\[
z_\nu(s)=\sqrt[\uproot{8}n_\mu]{c_\nu n_\nu
s+z_\nu(\bt(s))^{n_\mu}}.
\]
where
\begin{align*}
t_{\mu 1}(s)&:=t_{\mu 1}+(n_\mu+1)sc_\mu t_{\mu\, n_\mu+1},\\
&\quad\vdots\\
t_{\mu\, N_\mu-n_\mu}(s)&:= t_{\mu \, N_\mu-n_\mu}+N_\mu s c_\mu
t_{\mu\, N_\mu},\\
t_{\mu\, N_\mu-n_\mu+j}(s)&:= t_{\mu \, N_\mu-n_\mu+j},\quad j\geq
1.
\end{align*}
Integrating
\[
\pde{f_\mu}{s}+\beta_\mu(z_\mu)\pde{f_\mu}{z_\mu}=0
\]
we get
\[
f_\mu(z_\mu,s)=f_\mu\Big(\sqrt[\uproot{4}n_\mu]{-c_\mu n_\mu s
+z_\mu^{n_\mu}(s)}\Big)=f_\mu(z_\mu(\bt(s))).
\]

\subsection{Invariance conditions for additional symmetries and string equations}

We note from \eqref{addsymmgen} that the invariance condition under
an additional symmetry
\begin{equation}\label{sim3}
F_{\mu}^-=\sum_{\nu=0}^M(F_{\nu})_{(\nu,+)}-F_\mu=0,\quad \forall
\mu,
\end{equation}

Thus, all the functions $F_\mu$ must reduce to a unique function
$F_\mu=F\in\r$. Given a solution of the string equations
 \eqref{twistor} we may take
\[
F_\mu=P_\mu^{1+r}Q_\mu^{1+s},
\]
and conclude, that $F_\mu=F\in\r$, $\forall\mu$. Hence, string
equations determine solutions invariant under additional symmetries
characterized by the generators
\[
V_{\mu,rs}=P_\mu^{1+r}Q_\mu^{1+s},
\]
which close a Poisson algebra
\[
\{V_{\mu,rs},V_{\mu,r's'}\}=((r+1)(s'+1)-(r'+1)(s+1))V_{r+r'\,
s+s'}.
\]

In particular the functions $V_{r0}$   generate a Virasoro algebra.

\appendix
\section{Appendix: The right logarithmic  derivative}  Here we follow \cite{michor}. Given a manifold
$\mathcal T$, a Lie group $G$ with Lie algebra $\g$ and a map
$\psi:\mathcal T\rightarrow G$ we define the right logarithmic
derivative $\delta f\in\Omega^1(\mathcal T,\g)$ as the following
$\g$-valued 1-form
\[
\delta \psi(\xi)=T_{\psi(\bt)}(\mu^{\psi(\bt)^{-1}})\circ
T_{\bt}\psi(\xi)\quad \forall\xi\in T_{\bt} \mathcal T,\;
\bt\in\mathcal T,
\]
where $\mu^g(h)=g\cdot h$ is the left multiplication in the Lie
group. Recall that the right Maurer--Cartan form
$\kappa\in\Omega^1(G,\g)$ is a $\g$-valued 1-form over $G$ given
by
\[
\kappa_g:=T_g(\mu^{g^{-1}}),
\]
in terms of which
\[
\delta \psi=\psi^*\kappa.
\]

Given two maps $\psi,\phi:\mathcal T\rightarrow G$ then
\begin{equation}\label{rules}
\delta(\psi\cdot\phi)=\delta\psi+\Ad\psi(\delta\phi)
\end{equation}
and therefore
\[
\delta(\psi^{-1})=-\Ad\psi(\delta\psi).
\]

It also holds for $\omega:=\delta\psi$ and $z=\Ad\psi (Z)$ that
\begin{align}
\d\omega+\frac{1}{2}[\omega,\omega]=0,\label{zc}\\ \d
z=[\delta\psi,z]+\Ad\psi(\d Z).\label{dz}
\end{align}
If there is an exponential mapping $\exp:\g\rightarrow G$ we have
the formula
\[
T_X\exp(Y)=T_e\mu^{\exp X}\cdot \int_0^1\Ad(\exp(sX))Y\text{d}s.
\]
Thus, if $\psi=\exp X$ with $X:\mathcal T\rightarrow \g$ we have
\begin{align*}
\delta \psi(\xi)&=T_{\psi}\mu^{\psi^{-1}}(T_X\exp) T_{\bt}
X(\xi)=T_{\psi}\mu^{\psi^{-1}}\circ T_e\mu^{\psi}\cdot
\int_0^1\Ad(\exp(sX))( T_{\bt} X (\xi))\d s\\
&=\int_0^1\Ad(\exp(sX))( T_{\bt} X(\xi))\d s,\forall \xi\in
T_{\bt}\mathcal T,
\end{align*}
 that when we are allow to write $\Ad\exp X=\sum_{n=0}^\infty
(\ad X)^n/n!$ --for example if $G$ is a Banach--Lie group-- reads
\[
\delta \psi=\sum_{n=0}^\infty\frac{(\ad X)^n T_{\bt} X}{(n+1)!}.
\]

Given a smooth curve $X:\R\rightarrow\g$ we consider the problem
\begin{align*}
  \delta\psi (\partial_t)&= X(t),\quad \psi:\R\rightarrow G\\
  \psi(0)&=e.
\end{align*}
If there exists a solution is unique an local existence implies
global existence. We write $\text{evol}:C^\infty(\R,\g)\rightarrow
G$, with $\text{evol}(X(t))=g(1)$ and say, following Milnor, that
the Lie group is regular is evol exists and is smooth. That is
smooth curves in the Lie algebra integrates, in terms of the
 right logarithmic derivative, to smooth curves in the Lie group.

\section*{Acknowledgements}

Partial economical support from Direcci\'{o}n General de Ense\~{n}anza
Superior e Investigaci\'{o}n Cient\'{\i}fica n$^{\mbox{\scriptsize
\underline{o}}}$ BFM2002-01607, from European Science Foundation:
MISGAM and from Marie Curie FP6 RTN ENIGMA is acknowledge.


\begin{thebibliography}{99}


\bibitem{zab4} O. Agam, E. Bettelheim, P. Wiegmann and A. Zabrodin,
Phys. Rev. Lett. \textbf{88} (2002) 236801.

\bibitem{kod}  S. Aoyama and Y. Kodama, Commun. Math. Phys. \textbf{182},
 (1996) 185.

\bibitem{zab3} A. Boyarsky, A. Marsahakov, O. Ruchaysky,
P. Wiegmann and A. Zabrodin, Phys. Lett. B \textbf{515} (2001) 483.

 \bibitem{27} C. P. Boyer and J. D. Finley,  J. Math. Phys. {\bf 23} (1982)
1126.

\bibitem{gib}  Y. Gibbons and S.P. Tsarev, Phys. Lett  A \textbf{258},
(1999) 263


\bibitem{luis1} F. Guil, M. Ma\~{n}as and L. Mart\'{\i}nez Alonso, J. Phys. A: Math. \&
Gen.  \textbf{36} (2003) 4047.


\bibitem{39} F. Guil, M. Ma\~{n}as and L. Mart\'{\i}nez Alonso,  J. Phys. A: Math.\&  Gen. \textbf{36}
(2003) 6457.

\bibitem{kaz} V. Kazakov and A. Marsahakov, J. Phys. A: Math. \& Gen. \textbf{36} (2003)
3107.

\bibitem{michor} A. Kriegel, P. W.
Michor, \emph{The Convenient Setting of Global
  Analysis}, Mathematical Surveys and Monographs \textbf{53},
  American Mathematical Society (1997).


\bibitem{kon1} B. Konopelchenko, L. Martinez Alonso and O. Ragnisco, A : Math. \& Gen.
{\bf 34}, (2001) 10209.


\bibitem{krich1} I. M. Krichever, Func. Anal. Appl. \textbf{22} (1989)
200.


\bibitem{krichever} I. M. Krichever, Comm. Pure Appl. Math.
\textbf{47} (1992) 437.


\bibitem{zab5} I. Krichever, M. Mineev-Weinstein, P. Wiegmann and A. Zabrodin,
Physica D \textbf{198} (2004) 1.


\bibitem{h1}  B. A. Kuperschmidt and Yu. I.  Manin, Funk. Anal. Appl.  \textbf{11}, (3)
,31 (1977);  \textbf{17}, (1), 25 (1978).

\bibitem{manas1} M. Ma\~{n}as, \emph{The Principal Chiral Model as an Integrable System}
 in \emph{Harmonic Maps and Integrable
Systems}, A. P. Fordy \& J. C. Wood (editors), Aspects of
Mathematics \textbf{23} (1994) 147, Vieweg, Wiesbaden.

\bibitem{manas2} M. Ma\~{n}as, J. Phys. A: Math. Gen. {\bf 37}
(2004) 9195.

\bibitem{40}  L. Martinez Alonso and M. Ma\~{n}as, J. Math. Phys. \textbf{44} (2003) 3294.

\bibitem{mel2} L. Martinez Alonso and E. Medina, J. Phys. A: Math. \& Gen. \textbf{37} (2004)
12005.

\bibitem{mel3} L. Martinez Alonso and E. Medina, Phys. Lett. B \textbf{610} (2005) 227.

\bibitem{futuro} L. Martinez Alonso, E. Medina and M. Ma\~{n}as, \emph{String equations in the Whitham hierarchy:
 solutions and $\tau$-functions}, to appear.


\bibitem{zab2} M. Mineev-Weinstein, P. Wiegmann and A. Zabrodin,
Phys. Rev. Lett. \textbf{84} (2000) 5106.

\bibitem{semenov} A. G. Reiman and M. A. Semenov-Tyan-Shanskii, J.
Sov. Math \textbf{46} (1986) 1631.

\bibitem{24} K. Takasaki,  Commun. Math. Phys. \textbf{170} (1995) 101.

\bibitem{19}  K. Takasaki and T. Takebe,  Lett. Math.
Phys. \textbf{23} (1991) 205.

\bibitem{38} K. Takasaki and T. Takebe, Int. J. Mod. Phys. \textbf{A7} Suppl.1 (1992)
889.

\bibitem{20}  K. Takasaki and T. Takebe, Lett. Math.
Phys. \textbf{28} (1993) 165.

\bibitem{21}  K. Takasaki and T. Takebe,  Rev. Math.
Phys. \textbf{7} (1995) 743.

\bibitem{zab7} R. Teodorescu, E. Bettelheim, O. Agam, A. Zabrodin
and P. Wiegmann, Nuc. Phys. B \textbf{700} (2004) 521; Nuc. Phys. B
\textbf{704} (2005) 407.

\bibitem{zab1} P. W. Wiegmann and P. B. Zabrodin, Comm. Math.
Phys. \textbf{213} (2000) 523.

\bibitem{zab6} A. Zabrodin, Teor. Mat. Fiz.  \textbf{142} (2005)
197.

\bibitem{h2}  V. E. Zakharov, Func. Anal. Priloz. \textbf{14}, 89-98 (1980);
Physica \textbf{3D}, 193-202 (1981).

\bibitem{benney} V. E. Zakharov,
\emph{Dispersionless limit of integrable systems in 2 + 1 dimensions}, in \emph{Singular
Limits of Dispersive Waves},  N.M. Ercolani et al. (editors), Plenum
Press, NY, (1994) 165-174.

%
%
%



%
%
%
%
%

%
%
%
%
%
%
%
%
%
%
%
%
%
%
%
%
%

%
%
%
%
%

%

%
%
%
%
%
%
%
%
%
\end{thebibliography}
\end{document}